\documentclass[conference]{IEEEtran}
\IEEEoverridecommandlockouts
\usepackage{graphicx}
\usepackage{subcaption}
\usepackage{balance}
\usepackage{cite}

\usepackage[usenames,dvipsnames]{color}
\usepackage{colortbl}
\definecolor{mygray}{gray}{.8}

\usepackage{soul}
\usepackage{hyperref}
\usepackage{graphicx}
\usepackage{subcaption}
\usepackage{multirow}
\usepackage{listings}
\usepackage{comment}
\usepackage{fancybox}
\usepackage{paralist}
\usepackage{framed}
\usepackage{multicols}
\usepackage{color}
\usepackage{array}
\usepackage{amsmath}
\usepackage{xspace}
\usepackage{url}
\usepackage{soul}
\usepackage{hyperref}
\usepackage{multirow}
\usepackage{listings}
\usepackage{fancybox}
\usepackage{paralist}
\usepackage{ragged2e}
\usepackage{colortbl}
\usepackage{pgfplots}
\usepackage{pgfplotstable}
\usepackage[framemethod=TikZ]{mdframed}
\usepackage{algorithmic}
\usepackage{algorithm}

\usepackage{tcolorbox}
\makeatletter
\newcommand{\mybox}[1]{%
	\setbox0=\hbox{#1}%
	\setlength{\@tempdima}{\dimexpr\wd0+13pt}%
	\begin{tcolorbox}[boxrule=0.5pt, colback=white, arc=4pt,
		left=6pt,right=6pt,top=6pt,bottom=6pt,boxsep=0pt]
		#1
	\end{tcolorbox}
}

\pgfplotstableset{
	/color cells/min/.initial=0,
	/color cells/max/.initial=1000,
	/color cells/textcolor/.initial=,
	color cells/.code={%
		\pgfqkeys{/color cells}{#1}%
		\pgfkeysalso{%
			postproc cell content/.code={%
				\begingroup
				\pgfkeysgetvalue{/pgfplots/table/@preprocessed cell content}\value
				\ifx\value\empty
				\endgroup
				\else
				\pgfmathfloatparsenumber{\value}%
				\pgfmathfloattofixed{\pgfmathresult}%
				\let\value=\pgfmathresult
				\pgfplotscolormapaccess
				[\pgfkeysvalueof{/color cells/min}:\pgfkeysvalueof{/color cells/max}]%
				{\value}%
				{\pgfkeysvalueof{/pgfplots/colormap name}}%
				\pgfkeysgetvalue{/pgfplots/table/@cell content}\typesetvalue
				\pgfkeysgetvalue{/color cells/textcolor}\textcolorvalue
				\toks0=\expandafter{\typesetvalue}%
				\xdef\temp{%
					\noexpand\pgfkeysalso{%
						@cell content={%
							\noexpand\cellcolor[rgb]{\pgfmathresult}%
							\noexpand\definecolor{mapped color}{rgb}{\pgfmathresult}%
							\ifx\textcolorvalue\empty
							\else
							\noexpand\color{\textcolorvalue}%
							\fi
							\the\toks0 %
						}%
					}%
				}%
				\endgroup
				\temp
				\fi
			}%
		}%
	}
}

\usepackage{multicols}
\usepackage{color}
\usepackage{xcolor}
\usepackage{array}
\usepackage{amsmath}
\usepackage{centernot}
\usepackage{xspace}
\usepackage{url}
\usepackage{verbatim}
\usepackage{wrapfig}
\usepackage{tabularx}
\usepackage{tikz}
\usepackage{pgfplots}
\usepackage{pgf-pie}
\usetikzlibrary{pgfplots.statistics,calc}

\definecolor{songcolor}{RGB}{191,191,191}

\begin{document}

\title{Characterizing and Understanding Software Developer Networks in Security Development}
\author{\IEEEauthorblockN{Song Wang$^{*}$ and Nachi Nagappan$^{\$}$}\\
	\IEEEauthorblockA{
	$^*$York University; 
	$^\$$Microsoft Research\\
	wangsong@eecs.yorku.ca, nachin@microsoft.com}
}

\maketitle
\begin{abstract}
To build secure software, developers often work together during software development and maintenance to find, fix, and prevent security vulnerabilities. 
Examining the nature of developer interactions during their security activities regarding security introducing and fixing activities can provide insights for improving current practices.

In this work, we conduct a large-scale empirical study to characterize and understand developers' interactions during their security activities regarding security introducing and fixing, which involves more than 16K security fixing commits and over 28K security introducing commits from nine large-scale open-source software projects. 
For our analysis, we first examine whether a project is a hero-centric project when assessing developers' contribution in their security activities. 
Then we study the interaction patterns between developers, explore how the distribution of the patterns changes over time, 
and study the impact of developers' interactions on the quality of projects. 
In addition, we also characterize the nature of developer interaction in security activities in comparison to developer interaction in non-security activities (i.e., introducing and fixing non-security bugs). 

Among our findings we identify that: most of the experimental projects are non hero-centric projects when evaluating developers' contribution by using their security activities; 
there exist common dominating interaction patterns across our experimental projects; the distribution of interaction patterns has correlation with the quality of software projects. 
We believe the findings from this study can help developers understand how vulnerabilities originate and fix under the interactions of software developers.
\end{abstract}


\section{Introduction}
\label{sec:intro}
Building reliable and security software becomes more and more challenging in modern software development. 
As 
vulnerabilities can have catastrophic and irreversible impacts, e.g., the recent Heartbleed (CVE-2014-0160) cost more than US\$500 million to the global economy~\cite{Heartbleed}.

Developing secure software is a team effort,
developers work together to find, fix, and prevent security vulnerabilities and during which they form implicit collaborative developer networks~\cite{shin2011evaluating,meneely2009secure,meneely2011socio,meneely2010strengthening,meneely2013patch,zimmermann2010searching,meneely2011does,meneely2008predicting,sureka2011using,zimmermann2008predicting,trabelsi2015mining,bird2009putting,bhattacharya2012graph,younis2016assessing,wolf2009predicting,kumar2013evolution,zheng2008analyzing}. 
Understanding the structure of developer
interaction in security assurance practices can be helpful for building more secure software. 
Along this line, many developer network-related analyses have been proposed to deal with problems in real-world security practice such as vulnerabilities prediction~\cite{shin2011evaluating,zimmermann2010searching}, 
exploring the impact of human factors on security vulnerabilities~\cite{meneely2014empirical,meneely2009secure,meneely2010strengthening}, and monitoring vulnerabilities~\cite{trabelsi2015mining,sureka2011using}. 

\begin{figure}[!htb]
\centering
\includegraphics[width=0.4\textwidth]{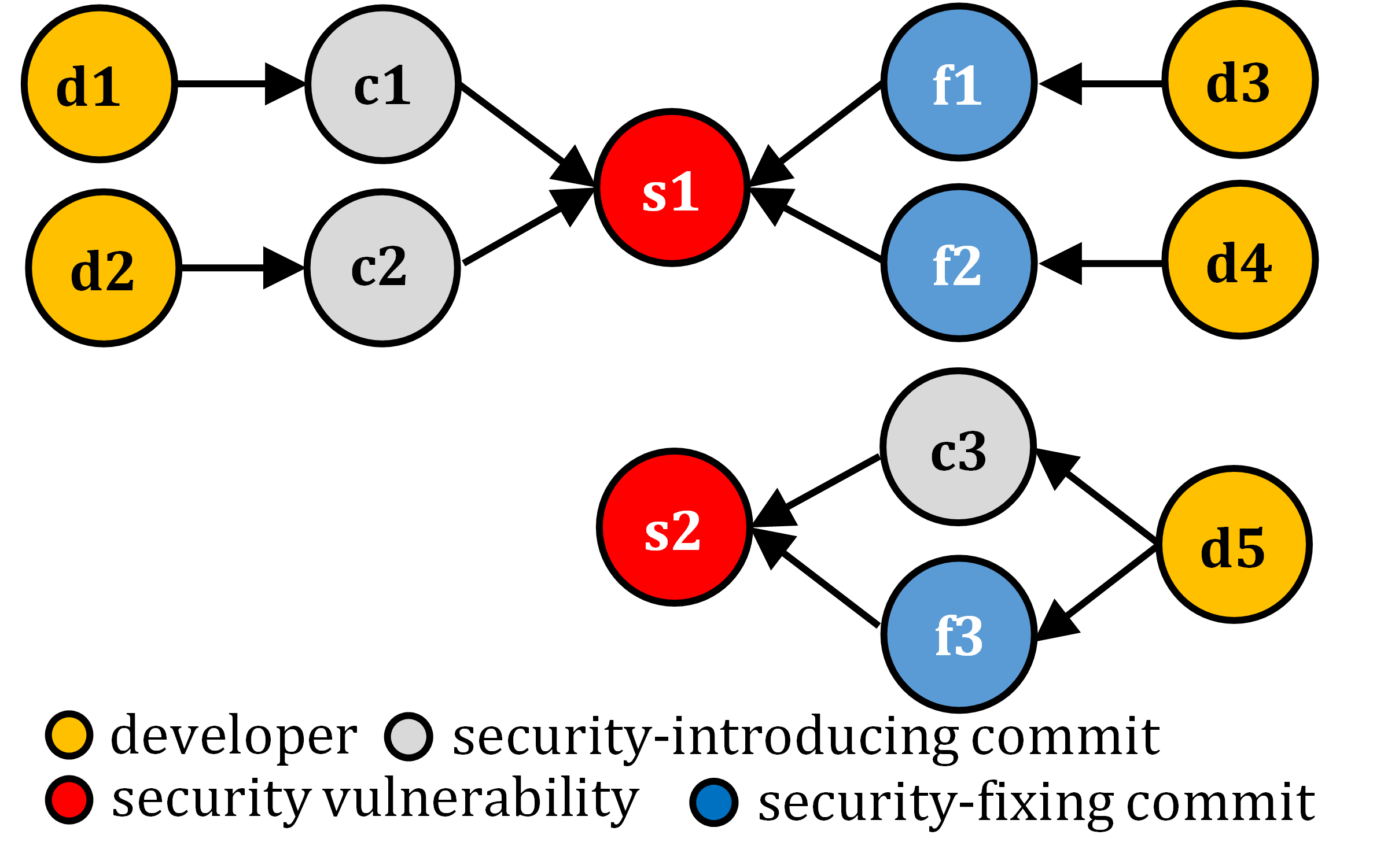}
\caption{An example developer security network.}
\label{fig:exmaple}
\vspace{-0.1in}
\end{figure}

Most of the existing approaches to exploring developer cooperation in security activities construct developer social network based on a single type of developer interaction, e.g., developers have co-changed/co-commented files that contain security vulnerabilities~\cite{shin2011evaluating,zimmermann2010searching,meneely2014empirical,meneely2009secure,meneely2010strengthening,trabelsi2015mining,sureka2011using}. 
However, security vulnerabilities are introduced and fixed by developers.  
During the life cycle of a security vulnerability, developers interact with each other via multiple ways. 
For example, as shown in Figure~\ref{fig:exmaple}, 
developers \texttt{d1} and \texttt{d2} introduced the security vulnerability \texttt{s1} via commit \texttt{c1} and \texttt{c2}; \texttt{s1} was later fixed by developer \texttt{d3} and  \texttt{d4} via commit \texttt{f1} and \texttt{f2}. 
The security vulnerability \texttt{s2}, was introduced via commit \texttt{c3} and fixed via commit \texttt{f3} by the same developer \texttt{d5}.  
Examining the nature of developer interactions during their security activities including both introducing and fixing security vulnerabilities can provide insights for improving current security practices.

In this paper, we propose the first study to characterize and understand developers' interactions in introducing and fixing security vulnerabilities by analyzing the developer networks built on their security activities. 
Our experiment dataset involves more than 16K security fixing commits and over 28K security introducing commits that ever appeared in nine large-scale open-source software projects including operation systems, compilers, PHP interpreter, Android platform, and JavaScript engine, etc. 
For our analysis, 
we first examine the heroism of software project when assessing developers' contribution in developers' security activities. 
As recent studies~\cite{agrawal2018we,majumder2019software,koch2002effort, mockus2002two,krishnamurthy2002cave,robles2009evolution} showed that most software projects are hero-centric projects where 80\% or more of the contributions (i.e., number of commits) are made by around 20\% of the developers.  
Then we explore whether there exist dominating interaction patterns between developers across our experimental projects, 
after that we study how the distribution of developers' interaction patterns changes in different projects over time. 
Finally we explore the potential impact of the developer interaction patterns on the quality of software projects by measuring the correlation between the changes of developer interactions and security density (i.e., dividing the number of security vulnerability by the number of submitted commits) in a given period of time. 
In addition, we also characterize the nature of developer interaction in security activities in comparison to developer interaction in non-security activities (i.e., introducing and fixing non-security bugs). 

This paper makes the following contributions:
\begin{itemize}
\item We conduct the first study to 
analyze developer interactions in security networks built on developers' security activities including both introducing and fixing security vulnerabilities.

\item We confirm that all experimental projects are hero-centric projects when assessing developers' contribution with non-security activities. 
However, we also find that most (eight out of nine) experimental projects are non hero-centric projects when assessing developers' contribution by using security activities.

\item We show that there exist dominating interaction patterns in both security and non-security activities across our experimental projects, while  
the distribution of developers' interaction in security and non-security activities are significantly different. 

\item We examine that developers' interaction is correlated  with the quality of a software project regarding security vulnerability density.
\end{itemize}	 

The rest of this paper is organized as follows. 
Section~\ref{sec:motivation} presents
the background.  
Section~\ref{sec:dataCollection} describes the
methodology of our approach. 
Section~\ref{sec:experiment} shows the experimental setup. 
Section~\ref{sec:result} presents the evaluation results. 
Section~\ref{sec:discussion} discusses the threats to the validity of this work. 
Section~\ref{sec:related} presents related studies. 
Section~\ref{sec:conclusion} concludes this paper. 
\section{Background}
\label{sec:motivation}
\subsection{Version-Control Systems}
\label{sec:vcs}
Version-control systems (VCS) are widely used in modern software development to coordinate developers' incremental contributions to a common software system. 
A VCS stores the entire source-code change history in the form of atomic change sets, called commits, which contain information about the changed code, the committers, and the timestamp of commits, etc.  
Git is one of the most popular VCSs, which has been adopted by more than 57M open-source projects and used by more than 20M developers\footnote{\url{https://en.wikipedia.org/wiki/GitHub}} globally.
Git's unique features make it especially appropriate for
mining invaluable information to better understand software process~\cite{bird2009promises,kalliamvakou2014promises}. 
For example, Git can track the history of lines as they are modified. 
By using the \texttt{git blame} feature, we can track the modification history of each line in a commit. 

In this work, we collect software security history data from nine projects that are maintained by Git to explore the developer interaction structures during their security activities, details are showed in Section~\ref{sec:dataCollection}.

\subsection{Developer Security Network}
\label{sec:dsc}
Developers interactions during their security activities including 
security fixing and introducing enable us to identify collaborative relationships between developers. 
The developer relationships can be described by a network, 
in which nodes represent developers and edges represent interactions between developers, in which nodes represent developers and edges represent interactions between developers.

In this study, a network can be formalized as a graph $G = (V, E)$, where $V$ is a set of vertices and $E$ is a set of edges, denoted by $V (G)$ and $E(G)$, respectively. 
An edge $e \in E$ is denoted as $e={v,u}$, where $v$ is the origin node and $u$ is the destination node from $V$. Graph edges are directed with different meanings. 

Different from most of existing developer social network studies~\cite{joblin2015developer,jermakovics2011mining,zhang2014developer,tymchuk2014collaboration,joblin2017classifying,ccaglayan2016effect,ren2018towards,palomba2018community,jermakovics2013exploring,joblin2017evolutionary,pinzger2008can,thung2013network,hong2011understanding,bird2008latent,surian2010mining,wang2013devnet,wolf2009mining,zhang2013heterogeneous,surian2011recommending,mcdonald2003recommending,jeong2009improving,zanetti2013categorizing,bird2006mining,zanetti2013rise,jiang2017mining,zhou2015will,zhou2012make,gharehyazie2015developer,lopez2006applying,lopez2004applying,toral2010analysis,canfora2011social,bird2011don,tsay2014influence,rahman2011ownership,zhou2015cross,izquierdo2011developers,german2003gnome}, 
in which $v \in V$ is a developer, and $e \in E$ represents a particular form of developer interactions, e.g., fixed bugs together~\cite{jeong2009improving,hong2011understanding,xuan2012developer,wang2013devnet}, co-changed files~\cite{pinzger2008can,wolf2009mining,shin2011evaluating,zimmermann2010searching,meneely2014empirical,meneely2009secure,meneely2010strengthening,trabelsi2015mining,sureka2011using}, 
worked on the same project~\cite{surian2011recommending}, or have communicated via email~\cite{bird2006mining}, etc., 
we consider a $v \in V$ in a developer security network may have three different types, 
i.e., developer, security-fixing commit, and security-introducing commit. 
Consequently, a $e \in E$ has also have three different types of meanings, i.e., 
a developer introduces a security vulnerability via a security-introducing commit, 
a developer fixes a security vulnerability via a security-fixing commit,
a security-fixing commit fixes the vulnerability introduced by a security-introducing commit.
\section{Data Collection Methodology}
\label{sec:dataCollection}
\begin{table*}[t!]
\centering
\caption{Experimental projects in this study. 
\textbf{Dev} is the number of developers.  
\textbf{Fix} is the number of commits that fixed security or non-security issues. 
\textbf{Intro} is the number of commits that 
introduced security or non-security issues.}

\label{tab:projects}
\setlength{\tabcolsep}{5pt}
\begin{tabular}{|l|l|l|l|l|l|l|l|l|l|l|l|}
\hline
\multirow{2}{*}{Project} & \multirow{2}{*}{Language} & \multirow{2}{*}{LastCommitDate} & \multirow{2}{*}{\#Commit} & \multirow{2}{*}{\#Dev} & \multirow{2}{*}{\#CVE} & \multicolumn{3}{c|}{Security Vulnerability}  & \multicolumn{3}{c|}{Non-Security Bugs}      \\ \cline{7-12} 
           &             &     &        &          &        & \multicolumn{1}{c|}{Fix} & \multicolumn{1}{c|}{Intro} & \multicolumn{1}{c|}{Dev} & \multicolumn{1}{c|}{Fix} & \multicolumn{1}{c|}{Intro} & \multicolumn{1}{c|}{Dev} \\ \hline
FFmpeg     & C/C++       & 2018/11/05        & 92,349 & 1,713    & 308    & 810        & 1,007        & 199 (11.62\%)      &16,024		&26,592		&1,138 (66.43\%)       \\ \hline
Freebsd    & C/C++       & 2018/11/05        & 255,969& 766      & 341    & 2,640      & 4,086        & 386 (50.39\%)      &35,776      &66,490		&604 (78.85\%)         \\ \hline
Gcc        & C/C++       & 2018/11/05        & 165,475& 604      & 6      & 575        & 1,341        & 200 (33.11\%)      &15,836      &29,975     &506 (83.77\%)       \\ \hline
Nodejs     & JS          & 2018/11/05        & 24,401 & 2,640    & 48     & 252        & 402          & 105 (3.98\%)       &4,792       &10,571  	&1,302 (49.32\%)        \\ \hline
Panda      & C/C++       & 2018/11/05        & 52,580 & 1,220    & 24     & 557        & 1,072        & 230 (18.85\%)      &9,133       &17,125 	&838 (68.69\%)         \\ \hline
Php        & C/C++       & 2018/11/05        & 109,461& 911      & 588    & 979        & 1,292        & 165 (18.11\%)      &25,610      &48,296     &663 (72.78\%)         \\ \hline
Qemu       & C/C++       & 2018/11/05        & 64,840 & 1,459    & 261    & 789        & 1,459        & 263 (18.03\%)      &12,139      &23,954     &1,023 (70.12\%)       \\ \hline
Linux      & C/C++       & 2018/11/05        & 796,003& 19,362   & 2,207  & 10,316     & 17,126       & 3,686 (19.04\%)    &174,687     &313,804    &14,046 (72.54\%)       \\ \hline
Android    & Java        & 2018/11/05        & 377,801& 2,938    & 1,763  & 2,439      & 2,521        & 496 (16.88\%)      &70,157		&128,893&2,132 (72.57\%)        \\ \hline
\end{tabular}
\vspace{-0.1in}
\end{table*}

\subsection{Subject Projects}
\label{subjects}
We selected nine open-source projects from existing studies~\cite{dinh2005freebsd,izurieta2006evolution,mockus2002two,tian2012identifying,joblin2015developer}, 
listed in Table~\ref{tab:projects}, to
explore developer interaction in security activities.  
The projects vary by the following dimensions: (a) size (lines of source code from
20 KLOC to over 17 MLOC, number of developers from 604
to 19K), (b) age (days since first commit), (c) programming language
(C/C++, Java, PHP, and JavaScript), (d) application domain
(operating system, compiler, PHP interpreter, Android platform, and JavaScript engine, etc.), 
and (e) VCS used (Git, Subversion). 
For each project, we extracted its code repository, and all the historical code commits hosted in GitHub on Nov. 5th 2018. 
Details of our approach to collecting the commits that introduce or fix security vulnerabilities and non-security bugs are as follows.

\subsection{Finding Public Vulnerabilities}
\label{sec:Vuln}

\subsubsection{Collecting Security Vulnerability Fixing Commits}
\label{sec:collectingSec}
Our data collection of security vulnerability fixing commits starts from the National Vulnerability Database (NVD)~\cite{nvd}, a database
provided by the U.S. National Institute of Standards and Technology
(NIST) with information pertaining to publicly disclosed software vulnerabilities. 
NVD contains entries for each publicly released vulnerability. 
These vulnerabilities are identified by CVE (Common Vulnerabilities and Exposures) IDs~\cite{cve}. 
When security researchers or vendors identify a vulnerability, they can request a CVE Numbering Authority to assign a CVE ID to it. 
Upon public release of the vulnerability information, 
the summarization the vulnerability, 
links to relevant external references (such as security fixing commits and issue reports), list of the affected software, etc., will be added to the CVEs. 
We first extracted all the public CVEs of each experimental  subject on Nov. 5th 2018.  
We then crawled the Git commit links to identify and clone the corresponding Git source code repositories and collected security
fixes using the commit hashes in the links. 
Note that, we also find that some of the external references only contain the bug/issue report links, 
e.g., the external reference of security vulnerability CVE-2018-14609\footnote{\url{https://nvd.nist.gov/vuln/detail/CVE-2018-14609}} does not contain the security fixing commits instead it shows the bug report ID\footnote{\url{https://bugzilla.kernel.org/show_bug.cgi?id=199833}}. 
For these security vulnerabilities, we used the fixing commits of these bugs as the security fixing commits. 
To collect the fixing commits of these bugs, we consider commits
whose commit messages contain the bug report ID as the fixing commits by following existing studies~\cite{kim2006automatic,tian2012identifying}. 


As reported in existing  studies~\cite{wijayasekara2012mining,ponta2019dataset}, not all security vulnerability have CVE identifiers, around 53\% of vulnerabilities in open source libraries are not disclosed publicly with CVEs~\cite{zhou2017automated,SourceClear}. 
To cover all possible vulnerabilities, 
we used the heuristical approaches proposed by Zhou et al.~\cite{zhou2017automated}, to identify the security fixing commits. 
Specifically, we used the regular expression rules listed in their Table 1, which included possible expressions and keywords related to security issues. 

\subsubsection{Grouping Security Fixing Commits}
\label{sec:grouping}
In the above section we have described how to collect security fixing commits. 
We found that some of the security fixing commits are made for fixing the same security vulnerability. 
For example, to fix security vulnerability CVE-2018-10883~\footnote{https://nvd.nist.gov/vuln/detail/CVE-2018-10883}, 
developers have made two commits.    
Identifying fixing commits that belong to the same security vulnerability could provide us valuable information about how vulnerabilities are fixed through developer interactions. 
To group fixing commits, 
first, for fixing commits that have CVE identifiers in their commit messages, we consider fixing commits that contain the same CVE identifiers belong to the same security vulnerabilities.   
Second, for fixing commits that do not have CVE identifiers in their commit messages, we propose a heuristical algorithm to group them, which is described in Algorithm~\ref{alg:algorithm1}. 
Specifically, given two fixing commits, we group them together  
if the similarity of their commit messages is larger than a threshold (i.e., $message_s$) and the modification location has overlaps. 
Following existing study~\cite{wang2019images,wang2016towards,runeson2007detection,rocha2016empirical}, we use the Cosine similarity to measure the similarity between two commit messages. 
We employ tf-idf~\cite{witten2016data}, stop words removal (e.g., ``is'', ``are'', and ``in'' since these words are used in most commit messages and thus have little discriminative power) and stemming (e.g., ``groups'' and ``grouping'' are reduced to ``group''.) to extract string vectors from the commit messages. 
For the threshold $thres_s$, we assume the ratios of 
collaborative fixing commits (i.e., fixing the same vulnerability) are similar between commits which have CVEs and commits that do not have CVEs. 
Thus for each project, we use the ratio of the collaborative fixing commits among the fixing commits that have CVEs to specify its threshold $thres_s$. 
In addition, we set the maximum interval between two collaborative fixing commits as six months, which is the typic length of fixing a security vulnerability~\cite{li2017large}.   
\begin{algorithm}[t]
	\footnotesize
	\caption{Grouping fixing commits algorithm}
	\label{alg:algorithm1}
	\hrule
	\begin{algorithmic}[1]
		\vspace{.2cm}
		\REQUIRE ~\\
		Fixing commit set $C$; \\
		Query fixing commit $q$;\\
		Commit message similarity threshold $thres_s$;\\
		Fixing location overlap rate threshold $modif_o$;\\
		\ENSURE ~\\
		A list of grouped fixing commit $D$;
		\FOR{each commit $r$ in $C$ and $q$}
		\STATE Extract commit messages and compute the similarity $message_s$; 
		\STATE Extract modified files and compute the overlap rate  $modif_o$;
		
		\IF{$message_s$ $\mathit{>}$ $thres_s$ and $modif_o$ $\mathit{>}$  $0$}  
		\STATE put $r$ in $\mathit{D}$
		\ENDIF
		\ENDFOR
	\end{algorithmic}
	\hrule
\end{algorithm}
\subsubsection{Collecting Security Vulnerability Introducing Commits}
\label{sec:security-intro}
With the above security-fixing commits, we further identify the security-introducing commits by using a blame technique
provided by a Version Control System (VCS), e.g., git or SZZ algorithm~\cite{kim2006automatic}. 
Following existing studies~\cite{perl2015vccfinder,sliwerski2005changes,da2017framework,jiang2013personalized}, 
we assume the deleted lines in a security-fixing commit are related to the root cause and considered as faulty lines. 
The most recent commit that introduced the faulty line is considered a  security-introducing commit. 

Note that, different from security-fixing commits, we did not group security-introducing commits. 
This is because the security-introducing commits are 
identified by security-fixing commits. 
Since we have already grouped security-fixing commits, 
these security-introducing commits are grouped accordingly. 
The details of the security-introducing commits as listed in Table~\ref{tab:projects}. 
The average number of security-introducing commits of a security-fixing commit ranges from 1.03 (Android) to 2.41 (Nodejs). 

\subsection{Finding Non-Security Bugs}
\label{sec:Non-Vuln}
\begin{figure}[t!]
\centering
\footnotesize
	\begin{tikzpicture}[thick, scale=0.8]
	\begin{axis}[
	ybar,
	height=4.2cm,
	width= 12cm,
	enlargelimits=0.1,
	legend style={at={(0.728, 0.93)}, anchor=north, legend columns=1},
	ymin=0,
	symbolic x coords={FFmpeg, Freebsd, Gcc, Nodejs, Panda, Php, Qemu, Linux, Android},
	xtick=data,
	x tick label style={font=\footnotesize},
	]
\addplot+[ybar,gray,  fill=gray]coordinates{(FFmpeg, 8.5) (Freebsd,2.4) (Gcc,6.7) (Nodejs,9.7) (Panda,9.2) (Php,11.3) (Qemu,5.8) (Linux,6.2) (Android,5.2)};
\addplot +[ybar, gray!25, fill=gray!25]coordinates{(FFmpeg, 9.6) (Freebsd,7.9) (Gcc,7.4) (Nodejs,11.2) (Panda,4.8) (Php,13.7) (Qemu,4.3) (Linux,9.6) (Android,6.5)};
\legend{Sec, Non-Sec}

\end{axis}
\end{tikzpicture}
\vspace{-0.05in}
\caption{\small The ratios (in percentage) of collaborative fixing commits grouped from security fixing and non-security fixing commits.}
\label{fig:groupRatio}
\vspace{-0.15in}
\end{figure}
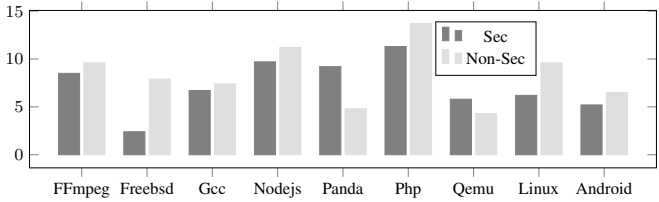

In addition, to explore the difference of developer interaction structures between developers' security activities and non-security activities, we also collect general bugs (i.e., non-security). 

Typically software bugs are discovered and reported to an issue
tracking system such as Bugzilla and later on fixed by the developers. A bug report usually records the description, the opening and 
fixing date, type (bug, enhancement, feature, etc.), etc. 
We consider a bug report in the Bugzilla database that is labelled as a ``bug'' to be a general bug. 
However, not all the projects have well-maintained bug tracking systems, 
in this work, following existing studies~\cite{sliwerski2005changes,da2017framework,jiang2013personalized}
if a project’s bug tracking system is not well maintained
and linked, we consider changes whose commit messages
contain the word ``fix'' and ``bug'' as bug-fixing commits.  
If a project’s bug tracking system is well maintained and linked, 
we consider commits whose commit messages contain a bug report ID as bug-fixing commits.
For each of the bug-fixing commit, we adopt the same approach as we used to identify security-introducing commits in Section~\ref{sec:security-intro}. 
Note that, if any of the non-security fixing commit appears in the security-fixing commit dataset, we will remove it from the non-security fixing commit dataset. 
The details of non-security fixing commits and their corresponding non-security introducing commits are showed in Table~\ref{tab:projects}. 
The average number of  non-security introducing commits of a non-security fixing commit ranges from 1.66 (FFmpeg) to 2.21 (Nodejs). 

In Section~\ref{sec:grouping}, we group security-fixing commits that fix the same security vulnerability. 
For non-security bugs, we also found the same phenomenon, i.e., some of the non-security fixing commits are made for fixing the same non-security bugs. 
For grouping these non-security fixing commits, we reuse Algorithm~\ref{alg:algorithm1}. 
As described in In Section~\ref{sec:grouping}, for grouping security fixing commits, we use the ratio of collaborative fixing commits (i.e., fix the same security vulnerability) that have CVE identifiers to set the threshold  $thres_s$ of a specific project.  
However, for non-security fixing commits, not all projects have well-maintained bug tracking systems, for some projects (e.g., Linux), we cannot use bug report ID to specify $thres_s$. 
Thus, we randomly pick and manually check 100 pairs of collaborative fixing commits on each the subject project, we use the average Cosine similarity value to set $thres_s$ in Algorithm~\ref{alg:algorithm1} to group non-security fixing commits.

Figure~\ref{fig:groupRatio} shows the ratios of collaborative fixing commits in the security fixing and non-security fixing commits. 
As we can see from the table, the ratios in non-security fixing commits are slightly higher than that of security fixing commits in most projects. 
On average, the ratio for security fixing commits is 7.2\% and the ratio for non-security fixing commits is 8.3\%, which is consistent with the finding from an existing study~\cite{gu2010has}, that 9\% of bug fixes were bad across three Java projects.

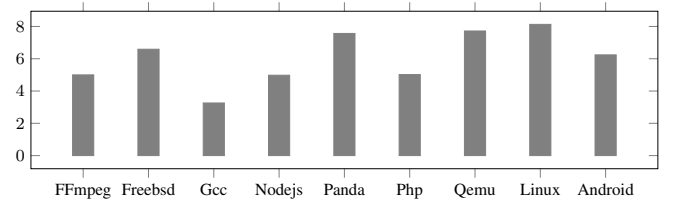
\begin{figure}[t!]
\centering
\footnotesize
	\begin{tikzpicture}[thick, scale=0.8]
	\begin{axis}[
	ybar,
	height=4.2cm,
	width= 12cm,
	enlargelimits=0.1,
	legend style={at={(0.728, 0.93)}, anchor=north, legend columns=1},
	ymin=0,
	symbolic x coords={FFmpeg, Freebsd, Gcc, Nodejs, Panda, Php, Qemu, Linux, Android},
	xtick=data,
	x tick label style={font=\footnotesize},
	]
\addplot+[ybar,gray,  fill=gray]coordinates{(FFmpeg, 5) (Freebsd, 6.59) (Gcc,3.26) (Nodejs, 4.98) (Panda, 7.567) (Php, 5.02) (Qemu, 7.72) (Linux, 8.13) (Android, 6.24)};

\end{axis}
\end{tikzpicture}
\vspace{-0.05in}
\caption{\small The overlap rate (in percentage) of non-security introducing commits and security introducing commits.}
\label{fig:overlap_introcommits}
\vspace{-0.15in}
\end{figure}

As we mentioned above, we have removed the non-security fixing commits from the security fixing commit dataset, while for non-security introducing commits and security introducing commits, 
we do not handle the overlaps, since it's possible that a 
security vulnerability and non-security bug can be introduced by the same introducing commit. 
In this work, we use overlap rate to measure the overlap level between two datasets. 
We define the \textbf{overlap rate} between datasets $A$ and $B$ as 
$\frac{A \cap B }{A \cup B}$. 
Figure~\ref{fig:overlap_introcommits} shows the overlap rates of non-security introducing commits and security introducing commits in the experimental projects. 
As we can see from the figure, the overlap rates of all experimental projects are lower than 10\%, which suggests that security vulnerability and non-security bugs usually have different introducing commits.

\subsection{Identifying Distinct Developers}
\label{sec:identifyingDev}
To build the developer security network, we need to obtain the developer information of security-fixing and security-introducing commits. 
In Git, for every pushed commit, Git maintains the user who did the commit, i.e., committer. Git computes the committer out of the Git configuration parameters `user.name' and `user.email'. Thus, by retrieving a commit, we can easily obtain its committer information. 
However, Git also allows users to change their profiles, 
which introduces the alias issue of developers in mining open-source~\cite{bird2006mining,robles2005developer}, 
i.e., a developer may have different emails/names. 
To solve this challenge, we use the aliases unmasking algorithms proposed in~\cite{bird2006mining} to identify distinct developers.
 
In total, we have around 45K distinct developers from the nine experimental projects, details are listed in Table~\ref{tab:projects}. 
Overall, the percentage of developer that involved in 
security activities ranges from 3.98\% to 50.39\%, while 
the percentage of developer that involved in 
non-security activities ranges from 49.32\% to 83.77\%.
\section{Research Question}
\label{sec:experiment}
Our experimental study is designed to answer the following
research questions. 

\vspace{4pt}
\textbf{RQ1. What are the distributions of developers in security and non-security activities?} 

Software security vulnerability and bugs are introduced and fixed by developers, in this RQ, we aim to explore the basic distribution of developers in security and non-security activities regarding fixing and introducing. 
For example, what is the overlap rate between developers that have ever involved in security activities and developers that have ever involved in non-security activities? 
What is the overlap rate between developers that have fixed security vulnerabilities and developers that have introduced security vulnerabilities?

\vspace{4pt}
\textbf{RQ2. How common are hero-centric projects regarding software security activities?}

Recent studies~\cite{agrawal2018we,majumder2019software,koch2002effort, mockus2002two,krishnamurthy2002cave,robles2009evolution} show that most software projects are hero-centric projects where 80\% or more of the contributions (e.g., the number of commits) are made by the 20\% of the developers. 
While the above studies 
assess developers' contribution from broad aspects, e.g., 
Agrawal et al.~\cite{agrawal2018we} used the number of commits made by each developer to represent its contribution to a project. 
Majumder et al.~\cite{majumder2019software} built a social interaction graph from developers' communication and used the node degree to represent a developer's contribution. 
Most of existing studies explore the heroism of projects from developers' code contribution and social communication perspectives. 
In this RQ, we aim to explore the heroism of a project when assessing developers' contribution by using a specific type of commits, e.g., security fixing, security introducing, non-security fixing, non-security introducing. 

\vspace{4pt}
\textbf{RQ3. What are the common interaction patterns between two developers in security activities?} 

Developers interact with each other during the development of a software project. 
In software development, the social and organizational aspects
have an impact on the individual and collective performance of the developers~\cite{ehrlich2012all}. Along this line, 
in this RQ, we aim to explore the common interaction structures among developers during their security and activities regarding security fixing and security introducing across different projects, 
which we believe can help us gain insight into distinct characteristics of developers' security activities. 

\vspace{4pt}
\textbf{RQ4. Are the distributions of developer interaction patterns in security and non-security activities different?} 

In this RQ, we aim to characterize the nature of developer interaction in security vulnerabilities in comparison to other non-security bugs. 
Specifically, we compare the distributions of interaction structures among developers in security vulnerabilities and non-security bugs. 

\vspace{4pt}
\textbf{RQ5. How do interaction structures among developers evolve over time?} 

Software team organization evolves over time~\cite{hong2011understanding,bird2008latent}, i.e., developers may leave a project and new developers may join during the life cycle of a project, 
which causes the evolution of developer community. 
Along this line, in this RQ, we aim to explore whether the interaction structure among developers changes over time and how it evolves during the life cycle of a project.

\vspace{4pt}
\textbf{RQ6. Does the change of interaction structures have an impact on the quality of software?} 

Developer social network and its evolution information have been examined could be used to predict new vulnerabilities and bugs~\cite{shin2011evaluating,zimmermann2010searching}. 
Along this line, in this RQ, we investigate whether the change of interaction structure has a correlation with the quality of software regarding the density of security vulnerabilities.
\section{Analysis Approach and Results}
\label{sec:result}
\subsection{\textbf{RQ1:} Distributions of Developers in Security and Non-Security Activities}
\label{sec:rq1}
\begin{table*}[th]
\centering
\caption{The overlap rates between developers that have been involved in different activities.
\textbf{secFix} denotes developers that have made security fixing commits,
\textbf{secIntro} denotes developers that have made security introducing commits,
\textbf{nonSecFix} denotes developers that have made non-security fixing commits,
\textbf{nonSecIntro} denotes developers that have made non-security introducing commits,
\textbf{sec} denotes developers that have made security fixing or introducing commits, and 
\textbf{nonSec} denotes developers that have made  non-security fixing or introducing commits. 
\textbf{secFix-secIntro} means the overlap rate between \textbf{secFix} and \textbf{secIntro}. 
The higher values with statistical significance ($p$-value $<$ 0.05) are shown with an asterisk (*). 
}

\label{tab:basic_overlap}
\begin{tabular}{|l|c|c|c|c|c|c|}
\hline
Project & secFix-secIntro (*) & secFix-nonSecFix & secFix-nonSecIntro & secIntro-nonSecFix & secIntro-nonSecIntro & sec-nonSec \\ \hline
FFmpeg  & 60.0            & 10.7             & 10.4               & 14.9               & 19.6                 & 10.7       \\ 
Freebsd & 89.0            & 30.2             & 32.2               & 49.1               & 43.2                 & 40.2       \\ 
Gcc     & 88.1            & 25.6             & 24.5               & 38.8               & 38.4                 & 25.6       \\ 
Nodejs    & 63.0            & 5.0              & 5.8                & 7.6                & 10.5                 & 5.0        \\
Panda   & 65.9            & 17.0             & 18.9               & 21.5               & 32.1                 & 17.0       \\
Php     & 70.5            & 15.5             & 16.9               & 21.7               & 29.3                 & 15.5       \\ 
Qemu    & 67.1            & 16.6             & 18.1               & 20.6               & 28.9                 & 16.6       \\
Linux   & 66.6            & 16.1             & 18.0               & 21.6               & 29.6                 & 16.1       \\
Android & 69.6            & 15.6             & 18.1               & 19.9               & 27.1                 & 15.6       \\ \hline  \hline
\textbf{Average} & 71.1            & 16.9             & 18.1               & 24.0               & 28.8                 & 18.0       \\ \hline

\end{tabular}
\vspace{-0.05in}
\end{table*}

To answer this RQ, we obtain unique developers from different activities, i.e., fixing security vulnerabilities, introducing security vulnerabilities, fixing non-security bugs, and introducing non-security bugs.  
Given the developer sets of two activities, we calculate their 
overlap rates via dividing the overlapping data points by all the unique data points.  
Table~\ref{tab:basic_overlap} shows the basic overlaps between developers that have been involved in different activities. 
As we can see from the table, in all the projects, developers from \textbf{secFix} and \textbf{secIntro} have higher overlap rates, i.e.,  
range from 60.0\% to 89.0\% and on average is 71.1\%, 
which indicates that most of the security vulnerabilities are introduced and fixed by a core group of developers. 
We can also see that the overlap rates of developers from security activities and non-security activities are lower, e.g., 
the overlap rate of 
developers from \textbf{secFix} and \textbf{nonSecFix} ranges from 5.0\% to 30.2\% and is 16.9\% on average, 
the overlap rate of developers from \textbf{secIntro} and \textbf{nonSecIntro} ranges from 19.6\% to 38.4\% and on average is 28.8\%. 
Overall, the overlap rate from \textbf{sec} and \textbf{nonSec} is 18.6\% on average, 
which indicates that most of the developers that are involved in security activities are different from developers that are involved in non-security activities. 
This may be because security issues are critical to software that require 
non-trivial domain expertise. Thus only a small group of developers 
is capable of handling security vulnerabilities, which makes the overlap rates of developers from security activities and non-security activities lower.
We further conduct the Wilcoxon signed-rank test ($p < 0.05$) to compare the overlap rates among different pairs. 
The results suggest that the overlap rates of \textbf{secFix} and \textbf{secIntro} are significantly higher than those of other pairs.

\mybox{Developers that are involved in security and non-security activities are different and have low overlap rates. For non-security activities, 
developers that introduced and fixed bugs have low overlap rates. 
However, for security activities, developers that introduced and fixed security vulnerabilities have higher overlap rates.} 
\subsection{\textbf{RQ2:} Heroism in Security and Non-Security Activities}
\label{sec:rq2}
\begin{table}[t]
\centering
\caption{The percentages of developers involved when contributing 80\% of a specific type of commits. 
Values with a red diamond (\textcolor{red}{$\diamond$}) indicate that a project is non hero-centric project.   
\textbf{All} denotes the combination of the four types of commits.
}
\label{tab:80_20dev}
\begin{tabular}{l|c|c|c|c|c}
\hline
Project & secFix & secIntro & nonSecFix & nonSecIntro& All \\ \hline
FFmpeg  & 20.1 (\textcolor{red}{$\diamond$}) & 17.1    & 3.6     & 5.5 & 3.5     \\ 
Freebsd & 32.1 (\textcolor{red}{$\diamond$})  & 26.0 (\textcolor{red}{$\diamond$})    & 13.4     & 11.3&  11.1     \\ 
Gcc     & 33.1 (\textcolor{red}{$\diamond$})  & 21.1 (\textcolor{red}{$\diamond$})    & 17.5     & 16.3&15.6       \\ 
Nodejs    & 34.0 (\textcolor{red}{$\diamond$})  & 24.7 (\textcolor{red}{$\diamond$})    & 13.5     & 6.6& 5.1      \\ 
Panda   & 36.5 (\textcolor{red}{$\diamond$})  & 25.8 (\textcolor{red}{$\diamond$})   & 10.6     &10.1& 7.5      \\ 
Php     & 21.6 (\textcolor{red}{$\diamond$})  & 23.7 (\textcolor{red}{$\diamond$})    & 6.3      & 8.2 & 5.7      \\ 
Qemu    & 30.1 (\textcolor{red}{$\diamond$}) & 22.3 (\textcolor{red}{$\diamond$})   & 8.6      & 9.8 & 6.8     \\ 
Linux   & 30.9 (\textcolor{red}{$\diamond$})  & 21.6 (\textcolor{red}{$\diamond$})   & 11.0     & 11.4& 8.5       \\ 
Android & 32.7 (\textcolor{red}{$\diamond$})  & 21.3 (\textcolor{red}{$\diamond$})    & 11.5     & 11.0 & 8.3     \\ \hline 
\end{tabular}
\vspace{-0.1in}
\end{table}
Following existing studies~\cite{agrawal2018we,majumder2019software}, 
we define a project to be hero-centric when 80\% of the contributions are done by about 20\% of the developers in this study. 
In addition, if the percentage of developers involved when contributing 80\% of 
a specific type of commits is larger than 20\%, 
we treat the project as non hero-centric projects. 

In this RQ, we first examine whether a project is a hero-centric project 
with only considering a specific type of commits, e.g,  
security fixing, security introducing, non-security fixing, and 
non-security introducing. 
To assess the contribution of a developer, 
following Agrawal et al.~\cite{agrawal2018we}, 
we count the number of a specific type of commits (i.e., 
security fixing, security introducing, non-security fixing, non-security introducing) made by each developer to 
represent his/her contribution to a project. 
We then rank developers ascendingly based on their contributions. 
Finally, we accumulate developers' contributions and record developers involved until 80\% of the contributions are done. 
In addition, we also evaluate a developer's contribution via the combination of the four types of commits.

Table~\ref{tab:80_20dev} shows the percentages of developers involved when contributing 80\% of a particular type of commits. 
As we can see from the table, all the projects are hero-centric projects, i.e., the percentages of developers involved are smaller than 20\%,  
when assessing developers' contribution 
by using non-security fixing or non-security introducing commits or all commits together. 
However, most of the experimental projects are non hero-centric projects when assessing developers' contribution by using security fixing or security introducing commits,  
e.g., the percentage of developers involved are 36.5\%, 
when evaluating developers' contribution by using security fixing commits in project Panda. 
Our finding indicates that although software development has ``heroes'', i.e., a small percentage of the staff who are responsible for most of the progress on a project, 
software security does not have typical ``heroes''. 

We further calculate the overlap rates of ``core developers'' (i.e., contribute 80\% of a specific type of commits) between different types of commits, which are shown in Table~\ref{tab:heor_overlap}. 
Note that we use ``core developers'' since a project can be a non hero-centric project  when assessing developers' contribution by using security activities. 
As we can see, the ``core developers'' from security fixing and security introducing have high overlap rates that range from 47.7\% to 71.1\% and on average is 63\%, which is consistent with our findings in Sec~\ref{sec:rq1}. 
The overlap rates of the ``core developers'' from security commits and non-security commits, 
i.e., ``core developers'' from \textbf{secFix} and \textbf{nonSecFix}, 
``core developers'' from \textbf{secIntro} and \textbf{nonSecFix} are 
lower than that of ``core developers'' only from security activities.  
This indicates that the ``core developers'' of security and non-security activities are different in most of the experimental projects.  

\mybox{All the experimental projects are examined as hero-centric projects in non-security activities, while most of them (8 out of 9) are non hero-centric projects in security activities.}
\begin{figure*}[tb]
    \centering
    \scriptsize
    \begin{subfigure}[b]{0.16\textwidth}
        \includegraphics[width=\textwidth]{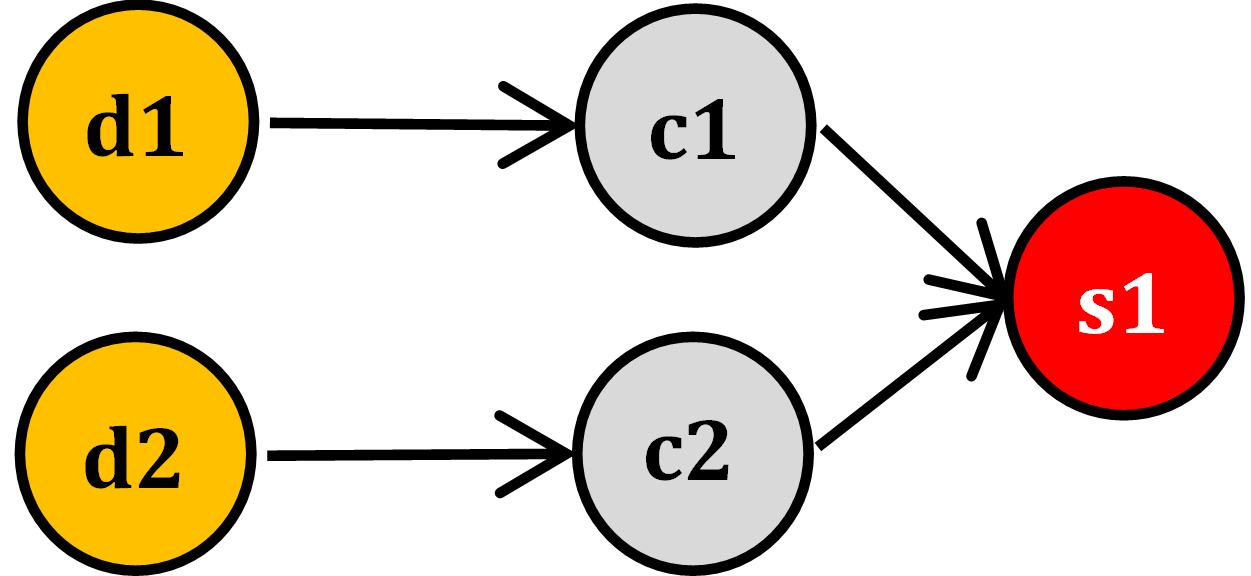}
        \caption{\texttt{\scriptsize P1: CoIntro}}
        \label{fig:r1}
    \end{subfigure}
    \begin{subfigure}[b]{0.16\textwidth}
        \includegraphics[width=\textwidth]{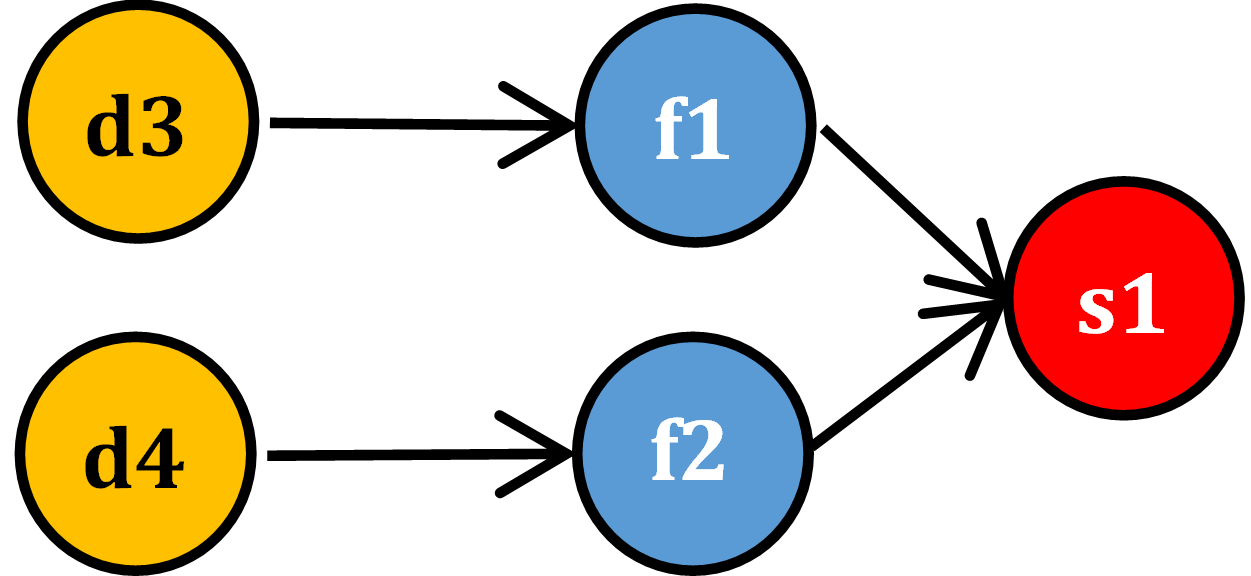}
        \caption{\texttt{\scriptsize P2: CoFix}}
        \label{fig:r2}
    \end{subfigure}
    \begin{subfigure}[b]{0.16\textwidth}
        \includegraphics[width=\textwidth]{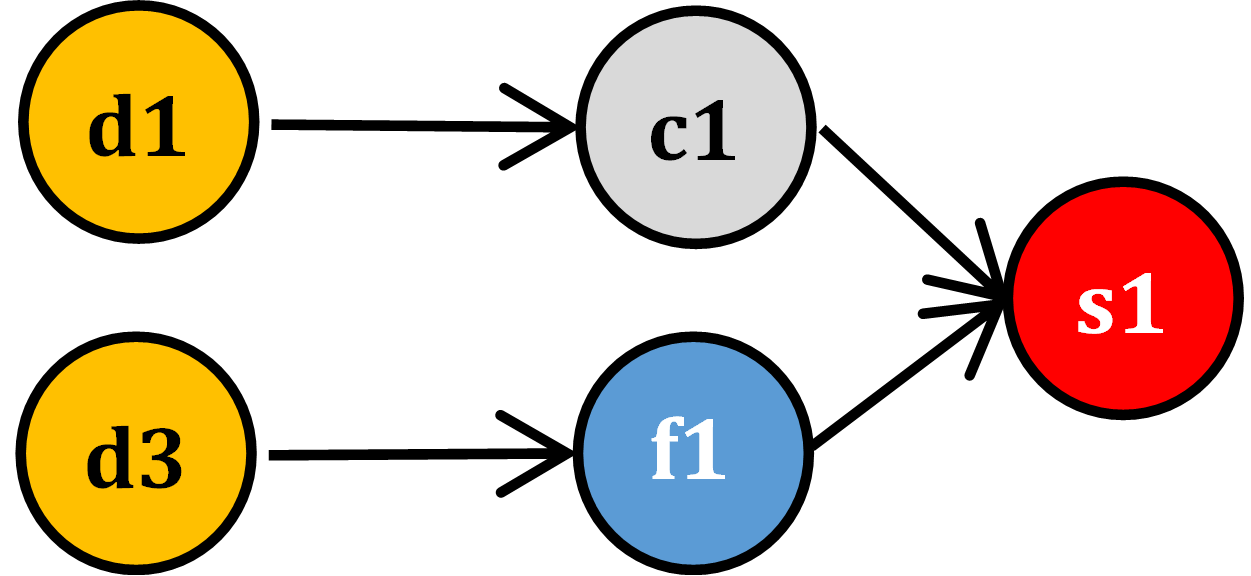}
        \caption{\texttt{\scriptsize P3: IntroFix}}
        \label{fig:r3}
    \end{subfigure}
	\begin{subfigure}[b]{0.16\textwidth}
        \includegraphics[width=\textwidth]{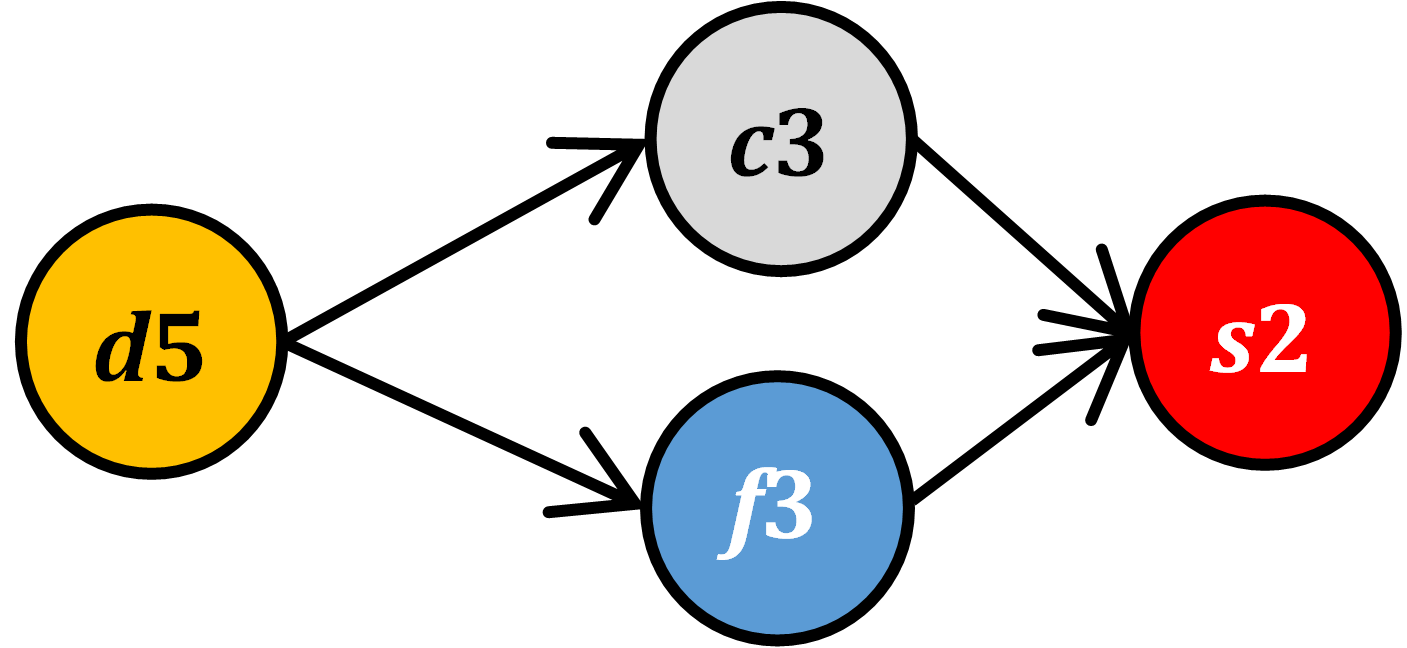}
        \caption{\texttt{\scriptsize P4: SelfIntroFix}}
        \label{fig:r4}
    \end{subfigure}
	\begin{subfigure}[b]{0.16\textwidth}
        \includegraphics[width=\textwidth]{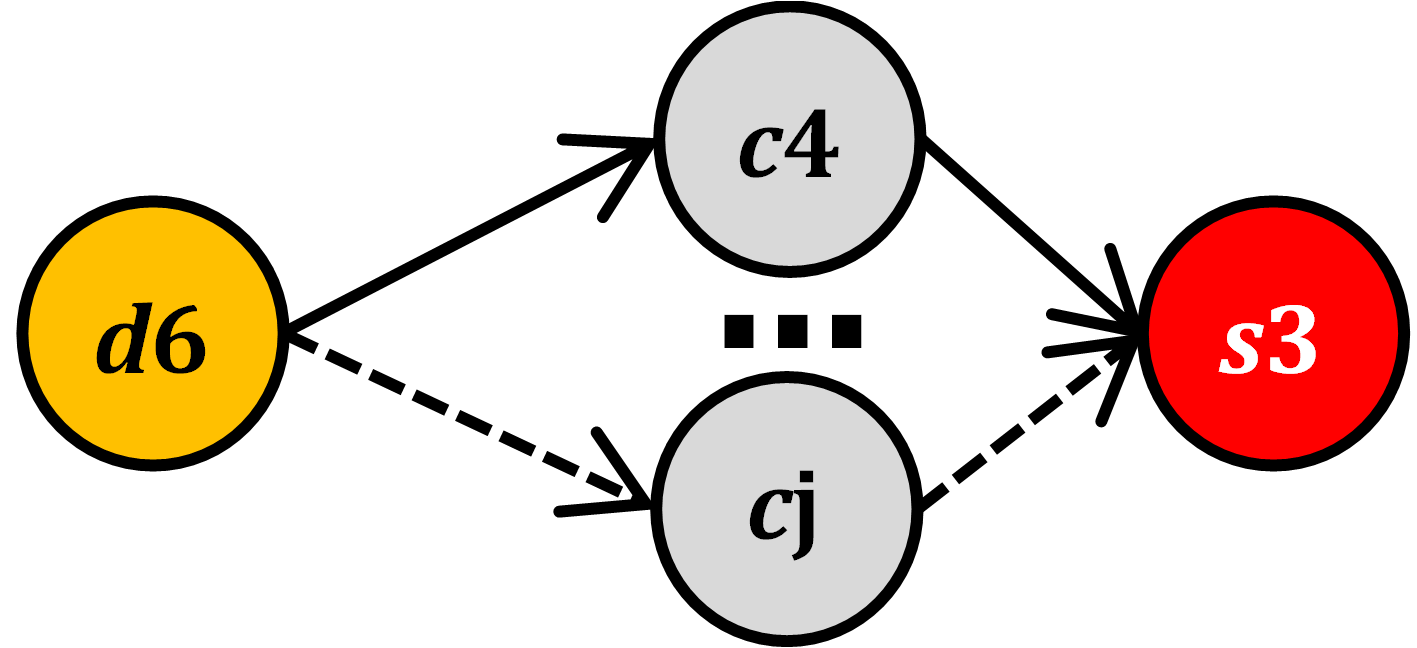}
        \caption{\texttt{\scriptsize P5: SelfIntro}}
        \label{fig:r5}
    \end{subfigure}
	\begin{subfigure}[b]{0.16\textwidth}
	\includegraphics[width=\textwidth]{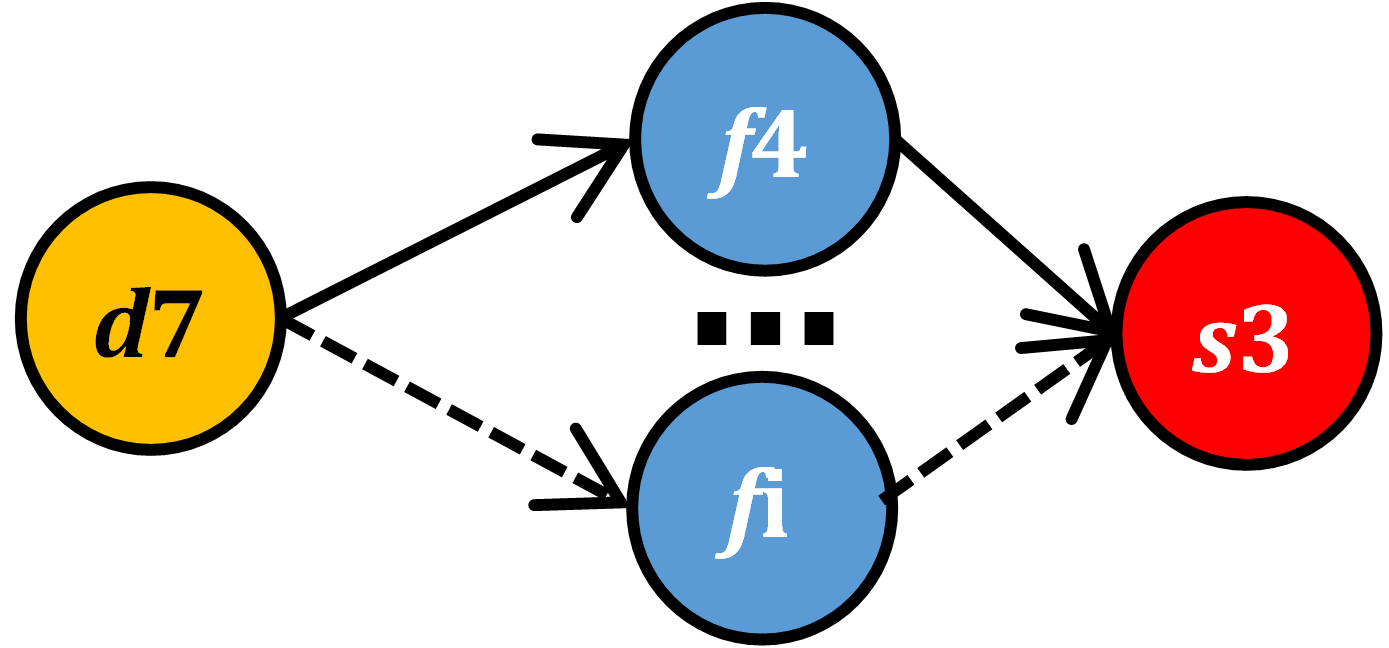}
	\caption{\texttt{\scriptsize P6: SelfFix}}
	\label{fig:r6}
\end{subfigure}
    \caption{The potential interaction relationships between developers during their security activities.}
	\label{fig:devRelationship}
	\vspace{-0.15in}
\end{figure*}
\begin{table}[tb]
\centering
\caption{The overlap rates of ``core developers'' (i.e., contribute 80\% of a particular type of commits) between different types of commits. 
The higher values with statistical significance ($p$-value $<$ 0.05) are shown with an asterisk (*). }
\label{tab:heor_overlap}
\setlength{\tabcolsep}{4pt}
\begin{tabular}{l|c|c|c}

\hline
Project & secFix-secIntro (*) & secFix-nonSecFix & secIntro-nonSecIntro \\ \hline
FFmpeg  & 71.1            & 50.0             & 68.9                \\ 
Freebsd & 69.9            & 55.6             & 67.8                 \\ 
Gcc     & 69.2            & 32.9             & 43.5                 \\ 
Nodejs    & 66.7            & 24.4             & 29.2                 \\ 
Panda   & 67.0            & 48.0             & 62.5                 \\ 
Php     & 48.5            & 54.3             & 66.7                 \\ 
Qemu    & 64.6            & 50.0             & 63.8                 \\
Linux   & 64.8            & 38.0             & 45.3                \\ 
Android & 47.0            & 34.3             & 45.8                 \\ \hline \hline
Average & 63.2            & 43.0             & 55.0   \\\hline
\end{tabular}

\end{table}
\subsection{\textbf{RQ3:} Common Developer Interaction Patterns in Developer Security Activities}
\label{sec:rq3}
\begin{table}[tb]
\centering
\caption{\small The distribution of developer interaction patterns during security activities (in percentage).}
\label{tab:secDist}
\setlength{\tabcolsep}{3pt}
\begin{tabular}{l|c|c|c|c|c|c}
\hline
Project & CoIntro & CoFix & IntroFix & SelfIntroFix & SelfIntro & SelfFix \\ \hline
FFmpeg  & 52.3    & 1.9   & 36.6     & 5.2          & 3.0       & 1.0     \\
Freebsd & 66.1    & 0.2   & 28.8     & 3.1          & 1.6       & 0.1     \\ 
Gcc     & 58.0    & 10.4  & 18.7     & 9.9          & 2.8       & 0.1     \\ 
Nodejs    & 50.9    & 7.4   & 23.4     & 10.7         & 7.3       & 0.3     \\ 
Panda   & 70.3    & 0.7   & 19.6     & 5.1          & 4.0       & 0.3     \\ 
Php     & 73.1    & 1.4   & 18.0     & 3.8          & 3.3       & 0.4     \\ 
Qemu    & 68.0    & 1.8   & 19.3     & 6.6          & 3.9       & 0.5     \\ 
Linux   & 67.1    & 0.5   & 21.9     & 5.8          & 4.5       & 0.2     \\ 
Android & 55.4    & 8.1   & 25.3     & 3.2          & 7.4       & 0.6     \\ \hline \hline
Average & 62.4    & 3.6   & 23.5     & 5.9        	& 4.2       & 0.4      \\ \hline
\end{tabular}
\vspace{-0.15in}
\end{table}

In this RQ, we identify developer interactions during the security activities including both introducing and fixing security vulnerabilities. 
Specifically, in order to explore developer interactions, 
we capture three possible interactions between two developers, i.e., two developers introduce the same security vulnerability (\texttt{CoIntro}), two developers fix the  same security vulnerability (\texttt{CoFix}), 
a security vulnerability is introduced by a developer and fixed by another developer (\texttt{IntroFix}), which are showed in Figure~\ref{fig:devRelationship} from \ref{fig:r1} to \ref{fig:r3}. 
In addition, we also collect the interactions of a single developer, i.e., a security vulnerability is introduced and fixed by a single developer (\texttt{SelfIntroFix}), 
a security vulnerability is introduced by multiple commits of a single developer and fixed by other developers (\texttt{SelfIntro}), and 
a security vulnerability is fixed by multiple commits of a single developer and is introduced by other developers (\texttt{SelfFix}), which are showed in Figure~\ref{fig:devRelationship} from \ref{fig:r4} to \ref{fig:r6}. 

For each subject project, we first build a security activity network, then with these interaction patterns, we further collect the numbers and calculate the percentages of the six patterns,  
which are showed in Table~\ref{tab:secDist}. 
As we can see from the figure, the six developer interaction patterns exist in each of the experimental projects. 
The \texttt{CoIntro} and \texttt{IntroFix} patterns are dominating (i.e., the accumulated percentage is larger than 80\%) across all the experimental projects. 
Other patterns take up around 20\% of developer interactions, 
for example, the percentages of\texttt{SelfFix} are lower than 1\% in all experimental projects. 
Although \texttt{CoIntro} and \texttt{IntroFix} are dominating, the percentages of them in different projects are different, i.e., range from 74.3\% (Nodejs) to 94.9\% (Freebsd).
In addition, the percentage of interactions between developers (i.e., \texttt{CoIntro},  \texttt{CoFix}, and  \texttt{IntroFix}) is much larger than that of 
interactions of the same developers (i.e., \texttt{SelfIntro},  \texttt{SelfFix}, and  \texttt{SelfIntroFix}), which indicates the nature of software security development is teamwork.

\mybox{The percentages of the developer interaction patterns vary dramatically in different projects. However, \texttt{CoIntro} and \texttt{IntroFix} patterns are dominating across all the experimental projects in developers' security activities.}

\subsection{\textbf{RQ4:} Comparison of Developer Interaction Patterns between Security and Non-Security Activities}
\label{sec:rq4}
In this RQ, we try to explore the difference of developer interactions between developers' security activities and non-security activities, which we believe can help us gain insight into distinct characteristics of developers' security activities. 
For each subject project, we first build a non-security activity network, then we further collect the ratios of the six patterns. 

Table~\ref{tab:disNonSec} shows the distribution of the six developer interaction patterns in developers' non-security activities. 
As we can see from the figure, 
although the six developer interaction patterns also exist in each of the experimental projects, the percentages of these patterns are different from that of security activities showed in Table~\ref{tab:secDist}. 
In Figure~\ref{fig:diffDist}, we show the detailed difference of interaction patterns between security activities and non-security activities. 
Specifically, the percentages of patterns \texttt{CoIntro} and \texttt{CoFix}, and \texttt{SelfIntro} vary dramatically across the experimental projects in this work. 

Different from security activities,
the dominating patterns (i.e., the accumulated percentage is larger than 80\%) in non-security activities include three patterns, i.e., \texttt{CoIntro}, \texttt{IntroFix}, and \texttt{CoFix}. 
Note that in security activities, the percentage of \texttt{CoFix} pattern ranges from 0.2\% to 10.4\% and on average is 3.5\%, while in non-security activities it ranges from 4.2\% to 26.1\% on average is 17.2\%. 
This may indicate that security vulnerability fixing requires domain expertise than fixing non-security bug fixing and most developers are incapable to fix security vulnerabilities, thus results in less teamwork. 

In addition, we also find that the dominating patterns are more balanced in developers' non-security activities compared to security activities. 
For example, the difference of the percentages of dominating patterns in security activities ranges from 15.7\% to 55.1\% and on average is 38.8\%, 
while in non-security activities, the difference ranges from 8.3\% to 35.2\% and on average is 21.2\%. 

\mybox{Developers have different dominating patterns in security and non-security activities. 
In addition, the distribution of developers' interaction in security and non-security activities are different.}
\begin{table}[tb]
\centering
\caption{The distribution of developer interaction during non-security activities (in percentage).}
\label{tab:disNonSec}
\setlength{\tabcolsep}{3pt}
\begin{tabular}{l|c|c|c|c|c|c}
\hline
Project & CoIntro & CoFix & IntroFix & SelfIntroFix & SelfIntro & SelfFix \\ \hline
FFmpeg  & 30.4    & 14.8  & 31.5     & 3.2          & 19.7      & 0.4     \\ 
Freebsd & 40.4    & 11.4  & 30.3     & 2.6          & 15.1      & 0.2     \\ 
Gcc     & 37.3    & 26.1  & 22.4     & 2.4          & 11.7      & 0.2     \\ 
Nodejs    & 59.0    & 26.3  & 9.7      & 1.5          & 3.4       & 0.1     \\ 
Panda   & 39.4    & 4.2   & 35.6     & 3.9          & 16.6      & 0.3     \\ 
Php     & 42.7    & 16.2  & 25.3     & 3.7          & 12.0      & 0.1     \\ 
Qemu    & 35.4    & 19.7  & 28.4     & 4.5          & 11.9      & 0.1     \\ 
Linux   & 32.1    & 15.5  & 33.0     & 3.5          & 15.8      & 0.1     \\ 
Android & 29.1    & 20.8  & 27.9     & 5.7          & 16.4      & 0.1     \\ \hline \hline
Average & 38.4    & 17.2  & 27.1     & 3.4          & 13.6      & 0.2    \\ \hline
\end{tabular}
\vspace{-0.1in}
\end{table}
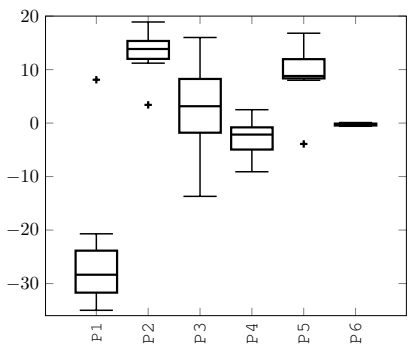
\begin{figure}[t!]
\centering
\begin{subfigure}{0.5\textwidth}
\centering
\scalebox{0.7}{
\begin{tikzpicture}
\pgfplotsset{
	   boxplot/every whisker/.style={thick,solid,black},
	   boxplot/every median/.style={very thick,solid,black},
}
\begin{axis}[
xtick={1,2,3,4,5,6},
xticklabels={ \small 
\texttt{P1},\texttt{P2},\texttt{P3},
\texttt{P4},\texttt{P5},\texttt{P6}
},
boxplot/draw direction=y,
ymin=-36,
ymax=20,
x tick label style={rotate=90,anchor=east},
]

\addplot[
very thick,
mark=+,
boxplot
]
table [row sep=\\, y index=0] {
	data\\-22.0\\-25.7\\-20.7\\8.1\\-30.9\\-30.4\\-32.5\\-35\\-26.3\\
};

\addplot[
very thick,
mark=+,
boxplot] 
table [row sep=\\, y index=0] {
data\\12.9\\11.2\\15.7\\18.9\\3.4\\14.8\\17.9\\15\\12.8\\
};

\addplot[
very thick,
mark=+,
boxplot]
table [row sep=\\, y index=0] {
data\\-5.0\\1.4\\3.7\\-13.7\\16\\7.3\\9.2\\11.1\\2.6\\
};

\addplot[
very thick,
mark=+,
boxplot]
table [row sep=\\, y index=0] {
data\\-2.1\\-0.5\\-7.6\\-9.1\\-1.1\\0\\-2.2\\-2.3\\2.5\\
};

\addplot[
very thick,
mark=+,
boxplot]
table [row sep=\\, y index=0] {
data\\16.8\\13.7\\8.7\\-3.9\\12.7\\8.7\\8\\11.2\\8.9\\
};

\addplot[
very thick,
mark=+,
boxplot]
table [row sep=\\, y index=0] {
data\\-0.6\\-0.1\\0.1\\-0.2\\-0.1\\-0.3\\-0.4\\-0.1\\-0.5\\
};

\end{axis}
\end{tikzpicture}
}
\end{subfigure}\hfill
\caption{ 
The difference of the percentages of interaction patterns between security activities and non-security activities. 
}
\label{fig:diffDist}
\end{figure}

\subsection{\textbf{RQ5:} Evolution of Developer Interaction in Developer Security Activities}
\label{sec:rq5}
\begin{figure*}[t!]

\begin{subfigure}{0.3\textwidth}
\centering
\scalebox{0.5}{
\begin{tikzpicture}
\pgfplotsset{
	   boxplot/every whisker/.style={thick,solid,black},
	   boxplot/every median/.style={very thick,solid,black},
}
\begin{axis}[
xtick={1,2,3,4,5,6},
xticklabels={ \small 
\texttt{P1},\texttt{P2},\texttt{P3},
\texttt{P4},\texttt{P5},\texttt{P6}
},
boxplot/draw direction=y,
ymin=-0.00,
ymax=0.8,
x tick label style={rotate=90,anchor=east},
]

\addplot[
very thick,
gray,
fill=gray,
mark=+,
boxplot
]
table [row sep=\\, y index=0] {
	data\\0.425287356\\0.630971993\\0.459219858\\0.489247312\\0.270967742\\0.357142857\\0.355191257\\0.227272727\\
};

\addplot[
very thick,
mark=+,
boxplot] 
table [row sep=\\, y index=0] {
data\\0.011494253\\0.004942339\\0.008865248\\0.010752688\\0.019354839\\0.007142857\\0.043715847\\0.02\\
};

\addplot[
very thick,
gray,
fill=gray,
mark=+,
boxplot]
table [row sep=\\, y index=0] {
data\\0.298850575\\0.220757825\\0.320921986\\0.266129032\\0.335483871\\0.332142857\\0.327868852\\0.381818182\\
};

\addplot[
very thick,
mark=+,
boxplot]
table [row sep=\\, y index=0] {
data\\0.096436782\\0.064250412\\0.060283688\\0.10483871\\0.2\\0.107142857\\0.087431694\\0.080909091\\
};

\addplot[
very thick,
gray,
fill=gray,
mark=+,
boxplot]
table [row sep=\\, y index=0] {
data\\0.046436782\\0.072487644\\0.045390071\\0.090967742\\0.061290323\\0.092857143\\0.080327869\\0.13818182\\
};

\addplot[
very thick,
mark=+,
boxplot]
table [row sep=\\, y index=0] {
data\\0.011494253\\0.006589786\\0.005319149\\0.008064516\\0.012903226\\0.003571429\\0.005464481\\0.018181818\\
};
\end{axis}
\end{tikzpicture}
}
\vspace{-0.0in}
\caption{FFmpeg}
\end{subfigure}\hfill
\begin{subfigure}{0.3\linewidth}
\centering
\scalebox{0.5}{
\begin{tikzpicture}
\pgfplotsset{
	   boxplot/every whisker/.style={thick,solid,black},
	   boxplot/every median/.style={very thick,solid,black},
}
\begin{axis}[
xtick={1,2,3,4,5,6},
xticklabels={ \small 
\texttt{P1},\texttt{P2},\texttt{P3},
\texttt{P4},\texttt{P5},\texttt{P6}
},
boxplot/draw direction=y,
ymin=-0.00,
ymax=0.8,
x tick label style={rotate=90,anchor=east},
]

\addplot[
very thick,
gray,
fill=gray,
mark=+,
boxplot
]
table [row sep=\\, y index=0] {
	data\\0.581352834\\0.52367688\\0.692810458\\0.629860031\\0.547307132\\0.5\\0.328621908\\0.537555228\\
};

\addplot[
very thick,
mark=+,
boxplot] 
table [row sep=\\, y index=0] {
data\\0\\0\\0.001633987\\0.00155521\\0.001455604\\0.004043127\\0.001766784\\0.004418262\\
};

\addplot[
very thick,
gray,
fill=gray,
mark=+,
boxplot]
table [row sep=\\, y index=0] {
data\\0.248628885\\0.300835655\\0.176470588\\0.231726283\\0.310043668\\0.318059299\\0.325088339\\0.288659794\\
};

\addplot[
very thick,
mark=+,
boxplot]
table [row sep=\\, y index=0] {
data\\0.115173675\\0.119777159\\0.080065359\\0.069984448\\0.07860262\\0.092991914\\0.204946996\\0.100147275\\
};

\addplot[
very thick,
gray,
fill=gray,
mark=+,
boxplot]
table [row sep=\\, y index=0] {
data\\0.054844607\\0.055710306\\0.047385621\\0.066874028\\0.062590975\\0.08490566\\0.136042403\\0.06921944\\
};

\addplot[
very thick,
mark=+,
boxplot]
table [row sep=\\, y index=0] {
data\\0.011494253\\0.006589786\\0.005319149\\0.008064516\\0.012903226\\0.003571429\\0.005464481\\0.018181818\\
};
\end{axis}
\end{tikzpicture}
}
\vspace{-0.0in}
\caption{Freebsd}
\end{subfigure}\hfill
\begin{subfigure}{0.3\linewidth}
\centering
\scalebox{0.5}{
\begin{tikzpicture}
\pgfplotsset{
	   boxplot/every whisker/.style={thick,solid,black},
	   boxplot/every median/.style={very thick,solid,black},
}
\begin{axis}[
xtick={1,2,3,4,5,6},
xticklabels={ \small 
\texttt{P1},\texttt{P2},\texttt{P3},
\texttt{P4},\texttt{P5},\texttt{P6}
},
boxplot/draw direction=y,
ymin=-0.00,
ymax=0.8,
x tick label style={rotate=90,anchor=east},
]

\addplot[
very thick,
gray,
fill=gray,
mark=+,
boxplot
]
table [row sep=\\, y index=0] {
	data\\0.653846154\\0.625\\0.493939394\\0.58326938\\0.51245552\\0.46645367\\0.48360656\\0.2365249\\
};

\addplot[
very thick,
mark=+,
boxplot] 
table [row sep=\\, y index=0] {
data\\0.159763314\\0.087797619\\0.065\\0.1202532617\\0.1108540925\\0.1032268371\\0.0918852459\\0.0954514442\\
};

\addplot[
very thick,
gray,
fill=gray,
mark=+,
boxplot]
table [row sep=\\, y index=0] {
data\\0.24260355\\0.25297619\\0.368181818\\0.121488872\\0.217081851\\0.2696486\\0.118852459\\0.13190812\\
};

\addplot[
very thick,
mark=+,
boxplot]
table [row sep=\\, y index=0] {
data\\0.162130178\\0.13125\\0.081818182\\0.14581734\\0.18505338\\0.05111821\\0.011065574\\0.25458267\\
};

\addplot[
very thick,
gray,
fill=gray,
mark=+,
boxplot]
table [row sep=\\, y index=0] {
data\\0\\0.0297619\\0.06060606\\0.03069839\\0.04626335\\0.01277955\\0.02868852\\0.03183989\\
};

\addplot[
very thick,
mark=+,
boxplot]
table [row sep=\\, y index=0] {
data\\0.011494253\\0.006589786\\0.005319149\\0.008064516\\0.012903226\\0.003571429\\0.005464481\\0.018181818\\
};
\end{axis}
\end{tikzpicture}
}
\vspace{-0.0in}
\caption{Gcc}
\end{subfigure}\hfill
\begin{subfigure}{0.3\linewidth}
\centering
\scalebox{0.5}{
\begin{tikzpicture}
\pgfplotsset{
	   boxplot/every whisker/.style={thick,solid,black},
	   boxplot/every median/.style={very thick,solid,black},
}
\begin{axis}[
xtick={1,2,3,4,5,6},
xticklabels={ \small 
\texttt{P1},\texttt{P2},\texttt{P3},
\texttt{P4},\texttt{P5},\texttt{P6}
},
boxplot/draw direction=y,
ymin=-0.00,
ymax=0.8,
x tick label style={rotate=90,anchor=east},
]

\addplot[
very thick,
gray,
fill=gray,
mark=+,
boxplot
]
table [row sep=\\, y index=0] {
	data\\0.25\\0.4\\0.363043478\\0.477310924\\0.444444444\\0.485412262\\0.3202486679\\0.527388535\\
};

\addplot[
very thick,
mark=+,
boxplot] 
table [row sep=\\, y index=0] {
data\\0.1\\0.09\\0.0904347826\\0.160504202\\0.177777778\\0.1945454545\\0.1055417407\\0.066390658\\
};

\addplot[
very thick,
gray,
fill=gray,
mark=+,
boxplot]
table [row sep=\\, y index=0] {
data\\0.375\\0.333333333\\0.206521739\\0.226890756\\0.333333333\\0.109936575\\0.076376554\\0.093418259\\
};

\addplot[
very thick,
mark=+,
boxplot]
table [row sep=\\, y index=0] {
data\\0.275\\0.266666667\\0.206521739\\0.176470588\\0.177777778\\0.135940803\\0.107300178\\0.111847134\\
};

\addplot[
very thick,
gray,
fill=gray,
mark=+,
boxplot]
table [row sep=\\, y index=0] {
data\\0\\0.12\\0.097826087\\0.058823529\\0.066666667\\0.093255814\\0.128419183\\0.078832272\\
};

\addplot[
very thick,
mark=+,
boxplot]
table [row sep=\\, y index=0] {
data\\0\\0\\0.02173913\\0\\0\\0\\0\\0.002123142\\
};
\end{axis}
\end{tikzpicture}
}
\vspace{-0.0in}
\caption{Nodejs}
\end{subfigure}\hfill
\begin{subfigure}{0.3\linewidth}
\centering
\scalebox{0.5}{
\begin{tikzpicture}
\pgfplotsset{
	   boxplot/every whisker/.style={thick,solid,black},
	   boxplot/every median/.style={very thick,solid,black},
}
\begin{axis}[
xtick={1,2,3,4,5,6},
xticklabels={ \small 
\texttt{P1},\texttt{P2},\texttt{P3},
\texttt{P4},\texttt{P5},\texttt{P6}
},
boxplot/draw direction=y,
ymin=-0.00,
ymax=0.8,
x tick label style={rotate=90,anchor=east},
]

\addplot[
very thick,
gray,
fill=gray,
mark=+,
boxplot
]
table [row sep=\\, y index=0] {
	data\\0.237623762\\0.603389831\\0.411255411\\0.521912351\\0.758752387\\0.584516129\\0.674418605\\0.911337209\\
};

\addplot[
very thick,
mark=+,
boxplot] 
table [row sep=\\, y index=0] {
data\\0.069306931\\0.003389831\\0.025974026\\0.011952191\\0.005092298\\0.00516129\\0.00620155\\0.003633721\\
};

\addplot[
very thick,
gray,
fill=gray,
mark=+,
boxplot]
table [row sep=\\, y index=0] {
data\\0.376237624\\0.311864407\\0.372294372\\0.282868526\\0.176957352\\0.268387097\\0.209302326\\0.068313953\\
};

\addplot[
very thick,
mark=+,
boxplot]
table [row sep=\\, y index=0] {
data\\0.148514851\\0.080508475\\0.073593074\\0.111553785\\0.063099936\\0.069677419\\0.075968992\\0.010174419\\
};

\addplot[
very thick,
gray,
fill=gray,
mark=+,
boxplot]
table [row sep=\\, y index=0] {
data\\0.158415842\\0.050847458\\0.116883117\\0.06374502\\0.026098027\\0.060645161\\0.03255814\\0.006540698\\
};

\addplot[
very thick,
mark=+,
boxplot]
table [row sep=\\, y index=0] {
data\\0.00990099\\0\\0\\0.007968127\\0\\0.011612903\\0.001550388\\0\\
};
\end{axis}
\end{tikzpicture}
}
\vspace{-0.0in}
\caption{Panda}
\end{subfigure}\hfill
\begin{subfigure}{0.3\linewidth}
\centering
\scalebox{0.5}{
\begin{tikzpicture}
\pgfplotsset{
	   boxplot/every whisker/.style={thick,solid,black},
	   boxplot/every median/.style={very thick,solid,black},
}
\begin{axis}[
xtick={1,2,3,4,5,6},
xticklabels={ \small 
\texttt{P1},\texttt{P2},\texttt{P3},
\texttt{P4},\texttt{P5},\texttt{P6}
},
boxplot/draw direction=y,
ymin=-0.00,
ymax=0.8,
x tick label style={rotate=90,anchor=east},
]

\addplot[
very thick,
gray,
fill=gray,
mark=+,
boxplot
]
table [row sep=\\, y index=0] {
	data\\0.313953488\\0.523219814\\0.844712182\\0.613475177\\0.696303696\\0.847280924\\0.758827949\\0.6159601\\
};

\addplot[
very thick,
mark=+,
boxplot] 
table [row sep=\\, y index=0] {
data\\0.023255814\\0.021671827\\0.004685408\\0\\0.012987013\\0.008171316\\0.023666416\\0.0174563591\\
};

\addplot[
very thick,
gray,
fill=gray,
mark=+,
boxplot]
table [row sep=\\, y index=0] {
data\\0.418604651\\0.294117647\\0.11914324\\0.237588652\\0.208791209\\0.113834883\\0.180315552\\0.229426434\\
};

\addplot[
very thick,
mark=+,
boxplot]
table [row sep=\\, y index=0] {
data\\0.127906977\\0.114551084\\0.007362784\\0.053191489\\0.046953047\\0.013243167\\0.019534185\\0.069825436\\
};

\addplot[
very thick,
gray,
fill=gray,
mark=+,
boxplot]
table [row sep=\\, y index=0] {
data\\0.11627907\\0.043343653\\0.023427041\\0.081560284\\0.026973027\\0.015497323\\0.014650639\\0.06234414\\
};

\addplot[
very thick,
mark=+,
boxplot]
table [row sep=\\, y index=0] {
data\\0\\0.003095975\\0.000669344\\0.014184397\\0.007992008\\0.001972387\\0.003005259\\0.004987531\\
};
\end{axis}
\end{tikzpicture}
}
\vspace{-0.0in}
\caption{Php}
\end{subfigure}\hfill
\begin{subfigure}{0.3\linewidth}
\centering
\scalebox{0.5}{
\begin{tikzpicture}
\pgfplotsset{
	   boxplot/every whisker/.style={thick,solid,black},
	   boxplot/every median/.style={very thick,solid,black},
}
\begin{axis}[
xtick={1,2,3,4,5,6},
xticklabels={ \small 
\texttt{P1},\texttt{P2},\texttt{P3},
\texttt{P4},\texttt{P5},\texttt{P6}
},
boxplot/draw direction=y,
ymin=-0.00,
ymax=0.8,
x tick label style={rotate=90,anchor=east},
]

\addplot[
very thick,
gray,
fill=gray,
mark=+,
boxplot
]
table [row sep=\\, y index=0] {
	data\\0.25\\0.601351351\\0.405982906\\0.524\\0.75443038\\0.574144487\\0.673374613\\0.767178363\\
};

\addplot[
very thick,
mark=+,
boxplot] 
table [row sep=\\, y index=0] {
data\\0.020833333\\0.006756757\\0.047008547\\0.016\\0.008860759\\0.030418251\\0.009287926\\0.021929825\\
};

\addplot[
very thick,
gray,
fill=gray,
mark=+,
boxplot]
table [row sep=\\, y index=0] {
data\\0.395833333\\0.310810811\\0.367521368\\0.284\\0.175949367\\0.263624842\\0.208978328\\0.12244152\\
};

\addplot[
very thick,
mark=+,
boxplot]
table [row sep=\\, y index=0] {
data\\0.15625\\0.080405405\\0.092649573\\0.112\\0.132911392\\0.098441065\\0.085851393\\0.0606725156\\
};

\addplot[
very thick,
gray,
fill=gray,
mark=+,
boxplot]
table [row sep=\\, y index=0] {
data\\0.166666667\\0.050675676\\0.106837607\\0.06\\0.025316456\\0.060836502\\0.03250774\\0.01754386\\
};

\addplot[
very thick,
mark=+,
boxplot]
table [row sep=\\, y index=0] {
data\\0.010416667\\0\\0\\0.004\\0.002531646\\0.002534854\\0\\0.010233918\\
};
\end{axis}
\end{tikzpicture}
}
\vspace{-0.0in}
\caption{Qemu}
\end{subfigure}\hfill
\begin{subfigure}{0.3\linewidth}
\centering
\scalebox{0.5}{
\begin{tikzpicture}
\pgfplotsset{
	   boxplot/every whisker/.style={thick,solid,black},
	   boxplot/every median/.style={very thick,solid,black},
}
\begin{axis}[
xtick={1,2,3,4,5,6},
xticklabels={ \small 
\texttt{P1},\texttt{P2},\texttt{P3},
\texttt{P4},\texttt{P5},\texttt{P6}
},
boxplot/draw direction=y,
ymin=-0.00,
ymax=0.8,
x tick label style={rotate=90,anchor=east},
]

\addplot[
very thick,
gray,
fill=gray,
mark=+,
boxplot
]
table [row sep=\\, y index=0] {
	data\\0.586815468\\0.646460537\\0.70097617\\0.646322599\\0.697541703\\0.651567498\\0.706114226\\0.716872485\\
};

\addplot[
very thick,
mark=+,
boxplot] 
table [row sep=\\, y index=0] {
data\\0.009343369\\0.0046786\\0.004306632\\0.003234426\\0.003621598\\0.003071017\\0.002627891\\0.002150687\\
};

\addplot[
very thick,
gray,
fill=gray,
mark=+,
boxplot]
table [row sep=\\, y index=0] {
data\\0.246301583\\0.227420667\\0.202842377\\0.240613135\\0.195456541\\0.243889955\\0.204011913\\0.203205217\\
};

\addplot[
very thick,
mark=+,
boxplot]
table [row sep=\\, y index=0] {
data\\0.094212302\\0.066924329\\0.053402239\\0.058079032\\0.061457419\\0.051823417\\0.046601261\\0.04016928\\
};

\addplot[
very thick,
gray,
fill=gray,
mark=+,
boxplot]
table [row sep=\\, y index=0] {
data\\0\\0.062808201\\0.053905614\\0.0374677\\0.051328927\\0.041812994\\0.049648113\\0.040119131\\0.037602331\\
};

\addplot[
very thick,
mark=+,
boxplot]
table [row sep=\\, y index=0] {
data\\0.000519076\\0.000610252\\0.001004881\\0.000421882\\0.000109745\\0\\0.000525578\\0\\
};
\end{axis}
\end{tikzpicture}
}
\vspace{-0.0in}
\caption{Linux}
\end{subfigure}\hfill
\begin{subfigure}{0.3\linewidth}
\centering
\scalebox{0.5}{
\begin{tikzpicture}
\pgfplotsset{
	   boxplot/every whisker/.style={thick,solid,black},
	   boxplot/every median/.style={very thick,solid,black},
}
\begin{axis}[
xtick={1,2,3,4,5,6},
xticklabels={ \small 
\texttt{P1},\texttt{P2},\texttt{P3},
\texttt{P4},\texttt{P5},\texttt{P6}
},
boxplot/draw direction=y,
ymin=-0.00,
ymax=0.8,
x tick label style={rotate=90,anchor=east},
]

\addplot[
very thick,
gray,
fill=gray,
mark=+,
boxplot
]
table [row sep=\\, y index=0] {
	data\\0.28600823\\0.384831461\\0.519490851\\0.418691589\\0.513278856\\0.482384824\\0.398836726\\0.57106383\\
};

\addplot[
very thick,
mark=+,
boxplot] 
table [row sep=\\, y index=0] {
data\\0.164691358\\0.164606742\\0.111137629\\0.126168224\\0.134729316\\0.126166817\\0.085999169\\0.085319149\\
};

\addplot[
very thick,
gray,
fill=gray,
mark=+,
boxplot]
table [row sep=\\, y index=0] {
data\\0.300411523\\0.183988764\\0.252187749\\0.293457944\\0.225740552\\0.20234869\\0.262982966\\0.188510638\\
};

\addplot[
very thick,
gray,
fill=gray,
mark=+,
boxplot]
table [row sep=\\, y index=0] {
data\\0.146090535\\0.110955056\\0.066030231\\0.140186916\\0.063329928\\0.048479374\\0.078105526\\0.042340426\\
};

\addplot[
very thick,
mark=+,
boxplot]
table [row sep=\\, y index=0] {
data\\0.234567901\\0.242977528\\0.14797136\\0.11588785\\0.155771195\\0.134296899\\0.165766514\\0.109787234\\
};

\addplot[
very thick,
mark=+,
boxplot]
table [row sep=\\, y index=0] {
data\\0.008230453\\0.012640449\\0.00318218\\0.005607477\\0.007150153\\0.006323397\\0.008309098\\0.002978723\\
};

\end{axis}
\end{tikzpicture}
}
\vspace{-0.0in}
\caption{Android}
\end{subfigure}\hfill

\caption{ 
The distribution of the number of different patterns in each project from 2009 to 2018. 
P1 denotes \texttt{CoIntro}, 
P2 denotes \texttt{CoFix},
P3 denotes \texttt{IntroFix},
P4 denotes \texttt{SelfIntroFix},
P5 denotes \texttt{SelfIntro}, and 
P6 denotes \texttt{SelfFix}.
}
\label{fig:rq5}
\vspace*{-0.0in}
\end{figure*}
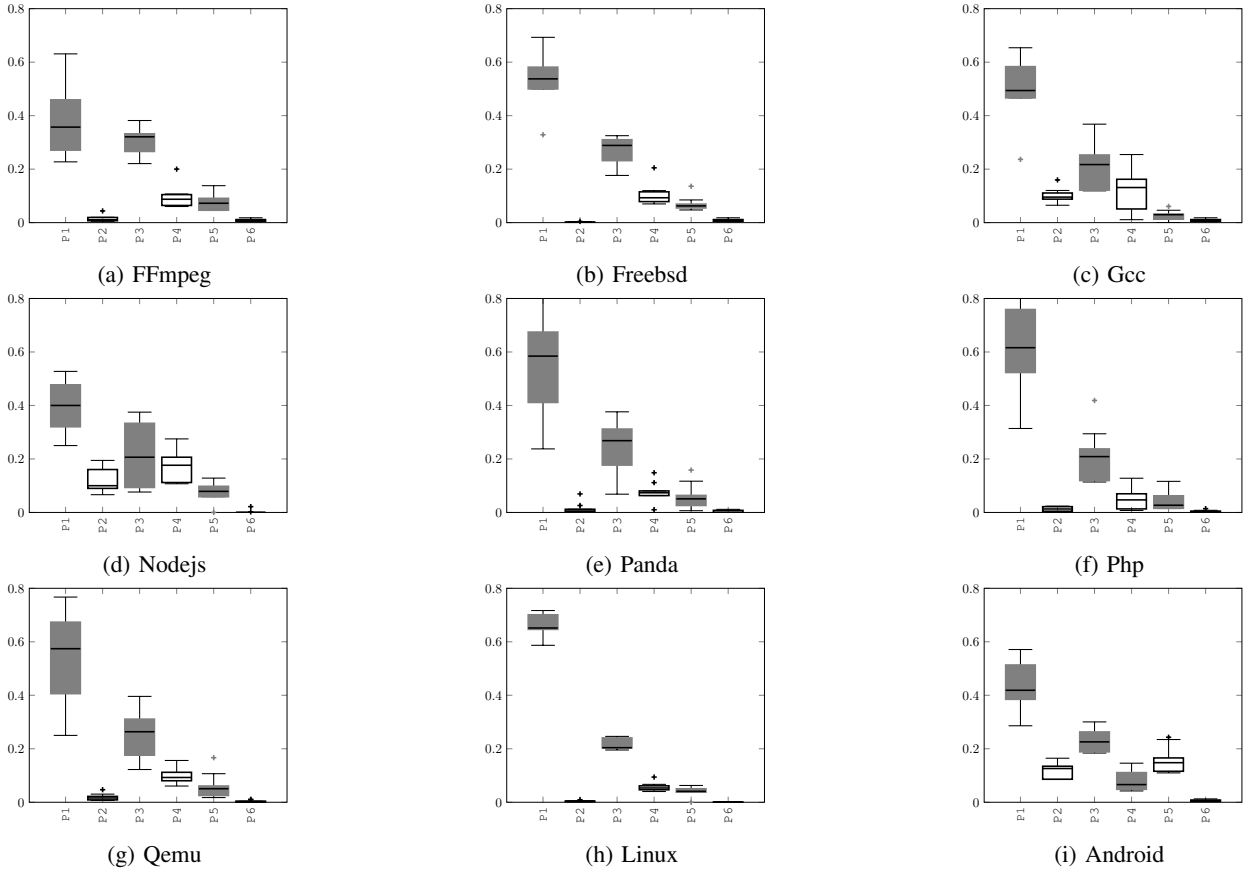

To explore the evolution of developer interactions, 
for each project, we collect the numbers and calculate the percentages of the six patterns that only appear in a specific year from 2009 to 2018.  
Thus, for each pattern, we have 10 different percentage values in each project. 
Figure~\ref{fig:rq5} shows the boxplots of the percentages of each interaction pattern in each project. 

The figure shows that overall, the percentages of a pattern vary dramatically in a project over time, for example, in FFmpeg, the percentages of pattern \texttt{CoIntro} range from 22.7\% to 63.1\% in 10 years. 
Regarding the dominating patterns, we find that patterns \texttt{CoIntro} and \texttt{IntroFix} are dominating on each project over time. 
In addition, the same phenomenon is also observed in developers' non-security activities. 

\mybox{The percentages of developer interaction patterns vary over time. 
While all the projects do not witness a change in terms of the dominating patterns.}

\subsection{\textbf{RQ6:} Impact of Developer Interaction on Software Quality}
\label{sec:rq6}
\begin{table}[tb]
\centering
\caption{The correlated patterns in each project.}
\label{tab:correlation}
\begin{tabular}{l|l}
\hline
Project &Correlated   Patterns\\ \hline
FFmpeg  & P1, P3, P5                                                      \\ 
Freebsd & P1, P3, P4                                                      \\ 
Gcc     & P1, P2, P3, P4                                                  \\ 
Nodejs    & P1, P2, P3, P4                                                  \\ 
Panda   & P1, P3                                                          \\ 
Php     & P1, P3, P4, P5                                                  \\ 
Qemu    & P1, P3, P5                                                      \\ 
Linux   & P1, P3                                                          \\ 
Android & P1, P2, P3, P4, P5                                              \\ \hline
\end{tabular}
\end{table}
To explore the relation between the changes of developers' interaction in security activity and the quality of the software, 
following existing studies~\cite{giger2012can,zimmermann2010searching}, 
we use the Spearman rank correlation~\cite{well2003research} to compute the correlations between the percentages of patterns and the density of security vulnerability appeared in each year from 2010 to 2018.  
The closer the value of a correlation is to +1 (or -1), the higher two measures are positively (or negatively).
A value of 0 indicates that two measures are independent.
Values greater than 0.10 can be considered a small effect size;
values greater than 0.30 can be considered a medium effect
size~\cite{zimmermann2010searching}. 
In this work, we consider the values larger than 0.10 
or smaller than -0.10 as correlated, others are uncorrelated.

Table~\ref{tab:correlation} shows the correlated patterns in each project. 
As we can see, five of the six patterns are selected as correlated in at least one project. In addition, the dominating patterns \texttt{CoIntro} and \texttt{IntroFix} are selected across all experimental projects. 

\mybox{Developers' interaction in security activities is correlated with the density of security vulnerabilities.}

\section{Threats to Validity}
\label{sec:discussion}

\paragraph{Internal Validity} 
Threats to internal validity are related to experimental errors.  
Following previous work~\cite{kim2006automatic,sliwerski2005changes,da2017framework,jiang2013personalized}, 
the process of collection security introducing or non-security introducing commits is 
automatically completed with the annotating or blaming function in VCS. 
It is known that this process can introduce
noise~\cite{kim2006automatic}. The noise in the data can potentially affect
the result of our study. 
Manual inspection of the process shows reasonable precision and recall on open
source projects~\cite{jiang2013personalized,kim2011dealing}. 
To mitigate this threat, we use the noise 
data filtering algorithm introduced in~\cite{kim2011dealing}.

\paragraph{External Validity} 
Threats to external validity are related
to the generalization of our study. 
The examined projects in this work have a large variance regarding project types. We have tried our best to make our dataset general and representative. 
However, it is still possible that the nine projects used in our experiments are not
generalizable enough to represent all software projects. 
Our approach might generate similar or different results for other projects that are not used in the experiments. 
We mitigate this threat by selecting projects of different functionalities
(operating systems, servers, and desktop applications) that
are developed in different programming languages (C, Java, and JavaScript).

In this work, all the experimental subjects are
open source projects. Although they are popular projects and widely used in security research, our findings may not be generalizable to commercial projects or
projects in other languages. 
\section{Related Work}
\label{sec:related}

\subsection{Developer Social Network}
\label{sec:network}
There has been a body of work that investigated aspects of developer social networks built on developers' activities during software  development~\cite{joblin2015developer,jermakovics2011mining,zhang2014developer,tymchuk2014collaboration,joblin2017classifying,ccaglayan2016effect,ren2018towards,palomba2018community,jermakovics2013exploring,joblin2017evolutionary,pinzger2008can,thung2013network,hong2011understanding,bird2008latent,surian2010mining,wang2013devnet,wolf2009mining,zhang2013heterogeneous,surian2011recommending,mcdonald2003recommending,jeong2009improving,zanetti2013categorizing,bird2006mining,zanetti2013rise,jiang2017mining,zhou2015will,zhou2012make,gharehyazie2015developer,lopez2006applying,lopez2004applying,toral2010analysis,canfora2011social,bird2011don,tsay2014influence,rahman2011ownership,zhou2015cross,izquierdo2011developers,german2003gnome,astromskis2017patterns,thung2013network}. 

Lopez-Fernandez et al.~\cite{lopez2006applying,lopez2004applying} first examined the social aspects of developer interaction during development, where developers were linked based on contributions to a common module. 
Bird et al.~\cite{bird2006mining} investigated developer organization and community structure in the mailing list of four open-source projects and used modularity as the community-significance measure to confirm the existence of statistically significant communities. 
Wolf et al.~\cite{wolf2009mining,wolf2009predicting} introduced an approach to mining developer collaboration from communication repositories and they further use developer collaboration to predict software build failures. 
Toral et al.~\cite{toral2010analysis} applied social-network analysis to investigate participation inequality in the Linux mailing list that contributes to role separation between core and peripheral contributors.
Hong et al.~\cite{hong2011understanding} and Zhang et al.~\cite{zhang2014developer} explored the characteristics of developer social networks built on developers interactions in bug tracking systems and how these networks evolve over time. 
Surian et al.~\cite{surian2010mining} extracted developer collaboration patterns from a large developer collaborations network extracted from SourceForge.Net, where developers are considered connected if both of them are listed as contributors to a project. 
Jeong et al.~\cite{jeong2009improving} and Xuan et al.~\cite{xuan2012developer} leveraged network metrics mined from social networks built in bug tracking systems to recommend developers for fixing new bugs. 
Surian et al.~\cite{surian2011recommending} used developer collaboration network extracted from Sourceforge.Net to recommend a list of top developers that are most compatible based on their programming language 
skills, past projects and project categories they have worked 
on before for a developer to work with. 
Researchers have also built social networks based on developers' security activities, i.e., have co-changed files that contain security vulnerabilities to predict new vulnerabilities~\cite{shin2011evaluating,zimmermann2010searching}, 
exploring the impact of human factors on security vulnerabilities~\cite{meneely2014empirical,meneely2009secure,meneely2010strengthening}, and monitoring vulnerabilities~\cite{trabelsi2015mining,sureka2011using}. 

Most of the above studies construct developer networks based on
a particular form of developer collaboration e.g.,
co-changed files, co-commented bugs, and co-contributed projects, etc., from bug tracking systems, mailing lists, or project contribution lists. 
These developer networks are homogeneous, which have merely one type of
node (developers) and one type of link (a particular form of developer collaboration). 
Wang et al.~\cite{wang2013devnet} and Zhang et al.~\cite{zhang2013heterogeneous} leveraged heterogeneous network analysis to mined different types of developer collaboration patterns in bug tracking system and further used these different collaborations to assist bug triage.

Our work differs in two ways from most of these prior studies: 
(1) We study developers' social interactions in security activities; 
(2) We explore different types of developer interactions during their security activities, which is more complex and with richer information.
\subsection{Security Vulnerability Analysis}
\label{sec:sec-analysis}
There are many studies to explore, analyze, and understand software security vulnerabilities~\cite{bu2017program,munaiah2018assisted,camilo2015bugs,decan2018impact,li2017large,frei2006large,shahzad2012large,zhong2015empirical,perl2015vccfinder,mu2018understanding,ozment2006milk,xu2017spain,walden2014predicting,medeiros2017software,yang2016mhcp}. 

Frei et al.~\cite{frei2006large} examined how vulnerabilities are handled with regard to information about discovery date, disclosure date, as well as the exploit and patch availability date in large-scale by analyzing more than 80,000 security advisories published between 1995 and 2006. 
Walden et al.~\cite{walden2014predicting} provided a  vulnerability dataset for evaluating the vulnerability prediction effectiveness of two modelling techniques, i.e., software metrics based and text mining based approaches. 
Medeiros et al.~\cite{medeiros2017software} examined the performance of software metrics on classifying vulnerable and non-vulnerable units of code. 
Yang et al.~\cite{yang2016mhcp} leveraged software network to evaluate structural characteristics of software systems during their evolution. 
Decan et al.~\cite{decan2018impact} and Shahzad~\cite{shahzad2012large} presented a large scale study of various aspects associated with software vulnerabilities during their life cycle. 
Ozment et al.~\cite{ozment2006milk} investigated the evolution of vulnerabilities in the OpenBSD operating system over time, observing that it took on average 2.6 years for a release version to remedy half of the known vulnerabilities. 
Perl et. al.~\cite{perl2015vccfinder} analyzed Git commits
that fixed vulnerabilities to produce a code analysis tool that assists in finding dangerous code commits. 
Xu et. al.~\cite{xu2017spain} developed a method for identifying security patches at the binary level based on execution traces, 
providing a method for obtaining and studying security patches on binaries and closed-source software. 
Li et al.~\cite{li2017large} conducted an analysis of various aspects of the patch development life cycle. 
There also existed some other studies that explored the characteristics of software general bugs~\cite{rahman2011ownership,zhou2015cross,izquierdo2011developers,german2003gnome,zhong2015empirical,park2012empirical}.

In this work, we propose the first study to characterize and understand developers' interaction by considering their activities in introducing and fixing security vulnerabilities by analyzing developer networks built on their security activities. 

\subsection{Heroism in Software Development}
\label{sec:heroism}
Heroism in software development is a widely studied topic.
Various researchers have found the presence of heroes in
software projects~\cite{agrawal2018we,majumder2019software,koch2002effort,
	mockus2002two,krishnamurthy2002cave,robles2009evolution}. 

Koch et al.~\cite{koch2002effort} studied the GNOME project 
and showed the presence of heroes throughout  the project history. 
Krishnamurthy~\cite{krishnamurthy2002cave} conducted a case study on 100 projects and 
reported that a few individuals are responsible for the main contribution of the projects. 
Agarwal et al.~\cite{agrawal2018we} studied heroism in software development on 661 open source projects from Github and 171 projects from an Enterprise Github.
They assess the contribution of a developer by the number of his/her commits submitted.   
Their experiment showed that 
77\% projects exhibit the pattern that 20\% of the total contributors complete 80\% of the contributions, which means hero-centric projects are very common in both 
public and enterprise projects. 
Majumder~\cite{majumder2019software} studies the heroes developer communities in 1100+ open source
GitHub projects. 
They built a social interaction graph from developers' communication and used the node degree to represent a developer's contribution. 
Based on the analysis, they found that hero-centric projects are majorly all projects. 

The above studies explore the heroism in software development from developers' code contribution and social communication perspectives. 
In this work, we examine whether software projects are hero-centric projects when assessing developers' contribution in their security activities. 
\section{Conclusion}
\label{sec:conclusion}
This work conducts a large-scale empirical study to characterize and understand developers' interaction during developers' security activities including both security vulnerability introducing and fixing activities, 
which involves more than 16K security fixing commits and over
28K security introducing commits from nine large-scale open-source software projects.
We first examine whether a project is a hero-centric project 
when assessing developers' contribution with developers' security activities.  
Then we examine the interaction patterns between developers in security activities,  
after that we show how the distribution of these patterns changes in different projects over time, 
finally we explore the potential impact of developers' interaction on the quality of projects by measuring the correlation between developers' interactions and 
the security density in a given period of time.  
In addition, we also characterize the nature of developer interaction in security activities in comparison to developer interaction in non-security activities (i.e., introducing and fixing non-security bugs). 

Among our findings we identify that: most of the experimental projects are non hero-centric projects when assessing developers' contribution by using security activities; different projects have different dominating interaction structures; developers' interaction has correlation with the quality of software projects. 
We believe the findings from this study can help developers understand how vulnerabilities originate and fix under the interaction of developers.


\bibliographystyle{IEEEtran}
\balance
\bibliography{paper}

\begin{thebibliography}{100}
\providecommand{\url}[1]{#1}
\csname url@samestyle\endcsname
\providecommand{\newblock}{\relax}
\providecommand{\bibinfo}[2]{#2}
\providecommand{\BIBentrySTDinterwordspacing}{\spaceskip=0pt\relax}
\providecommand{\BIBentryALTinterwordstretchfactor}{4}
\providecommand{\BIBentryALTinterwordspacing}{\spaceskip=\fontdimen2\font plus
\BIBentryALTinterwordstretchfactor\fontdimen3\font minus
  \fontdimen4\font\relax}
\providecommand{\BIBforeignlanguage}[2]{{%
\expandafter\ifx\csname l@#1\endcsname\relax
\typeout{** WARNING: IEEEtran.bst: No hyphenation pattern has been}%
\typeout{** loaded for the language `#1'. Using the pattern for}%
\typeout{** the default language instead.}%
\else
\language=\csname l@#1\endcsname
\fi
#2}}
\providecommand{\BIBdecl}{\relax}
\BIBdecl

\bibitem{Heartbleed}
``Heartbleed,'' \url{http://heartbleed.com/}.

\bibitem{shin2011evaluating}
Y.~Shin, A.~Meneely, L.~Williams, and J.~A. Osborne, ``Evaluating complexity,
  code churn, and developer activity metrics as indicators of software
  vulnerabilities,'' \emph{TSE'11}, vol.~37, no.~6, pp. 772--787.

\bibitem{meneely2009secure}
A.~Meneely and L.~Williams, ``Secure open source collaboration: an empirical
  study of linus' law,'' in \emph{CCS'09}, pp. 453--462.

\bibitem{meneely2011socio}
------, ``Socio-technical developer networks: Should we trust our
  measurements?'' in \emph{ICSE'11}, pp. 281--290.

\bibitem{meneely2010strengthening}
------, ``Strengthening the empirical analysis of the relationship between
  linus' law and software security,'' in \emph{ESEM'10}, p.~9.

\bibitem{meneely2013patch}
A.~Meneely, H.~Srinivasan, A.~Musa, A.~R. Tejeda, M.~Mokary, and B.~Spates,
  ``When a patch goes bad: Exploring the properties of
  vulnerability-contributing commits,'' in \emph{ESEM'13}, pp. 65--74.

\bibitem{zimmermann2010searching}
T.~Zimmermann, N.~Nagappan, and L.~Williams, ``Searching for a needle in a
  haystack: Predicting security vulnerabilities for windows vista,'' in
  \emph{ICST'10}, pp. 421--428.

\bibitem{meneely2011does}
A.~Meneely, P.~Rotella, and L.~Williams, ``Does adding manpower also affect
  quality?: an empirical, longitudinal analysis,'' in \emph{FSE'11}, pp.
  81--90.

\bibitem{meneely2008predicting}
A.~Meneely, L.~Williams, W.~Snipes, and J.~Osborne, ``Predicting failures with
  developer networks and social network analysis,'' in \emph{FSE'08}, pp.
  13--23.

\bibitem{sureka2011using}
A.~Sureka, A.~Goyal, and A.~Rastogi, ``Using social network analysis for mining
  collaboration data in a defect tracking system for risk and vulnerability
  analysis,'' in \emph{ISEC'11}, pp. 195--204.

\bibitem{zimmermann2008predicting}
T.~Zimmermann and N.~Nagappan, ``Predicting defects using network analysis on
  dependency graphs,'' in \emph{ICSE'08}, pp. 531--540.

\bibitem{trabelsi2015mining}
S.~Trabelsi, H.~Plate, A.~Abida, M.~M.~B. Aoun, A.~Zouaoui, C.~Missaoui,
  S.~Gharbi, and A.~Ayari, ``Mining social networks for software
  vulnerabilities monitoring,'' in \emph{NTMS'15}, pp. 1--7.

\bibitem{bird2009putting}
C.~Bird, N.~Nagappan, H.~Gall, B.~Murphy, and P.~Devanbu, ``Putting it all
  together: Using socio-technical networks to predict failures,'' in
  \emph{ISSRE'09}, pp. 109--119.

\bibitem{bhattacharya2012graph}
P.~Bhattacharya, M.~Iliofotou, I.~Neamtiu, and M.~Faloutsos, ``Graph-based
  analysis and prediction for software evolution,'' in \emph{ICSE'12}, pp.
  419--429.

\bibitem{younis2016assessing}
A.~Younis, Y.~K. Malaiya, and I.~Ray, ``Assessing vulnerability exploitability
  risk using software properties,'' \emph{SQJ'16}, vol.~24, no.~1, pp.
  159--202.

\bibitem{wolf2009predicting}
T.~Wolf, A.~Schroter, D.~Damian, and T.~Nguyen, ``Predicting build failures
  using social network analysis on developer communication,'' in
  \emph{ICSE'09}, pp. 1--11.

\bibitem{kumar2013evolution}
A.~Kumar and A.~Gupta, ``Evolution of developer social network and its impact
  on bug fixing process,'' in \emph{ISEC'13}, pp. 63--72.

\bibitem{zheng2008analyzing}
X.~Zheng, D.~Zeng, H.~Li, and F.~Wang, ``Analyzing open-source software systems
  as complex networks,'' \emph{Physica A'08}, vol. 387, no.~24, pp. 6190--6200.

\bibitem{meneely2014empirical}
A.~Meneely, A.~C.~R. Tejeda, B.~Spates, S.~Trudeau, D.~Neuberger, K.~Whitlock,
  C.~Ketant, and K.~Davis, ``An empirical investigation of socio-technical code
  review metrics and security vulnerabilities,'' in \emph{SSE'14}, pp. 37--44.

\bibitem{agrawal2018we}
A.~Agrawal, A.~Rahman, R.~Krishna, A.~Sobran, and T.~Menzies, ``We don't need
  another hero?: the impact of heroes on software development,'' in
  \emph{ICSE-SEIP'18}, pp. 245--253.

\bibitem{majumder2019software}
S.~Majumder, J.~Chakraborty, A.~Agrawal, and T.~Menzies, ``Why software
  projects need heroes (lessons learned from 1100+ projects),'' \emph{arXiv
  preprint arXiv:1904.09954}, 2019.

\bibitem{koch2002effort}
S.~Koch and G.~Schneider, ``Effort, co-operation and co-ordination in an open
  source software project: Gnome,'' \emph{Information Systems Journal'02},
  vol.~12, no.~1, pp. 27--42.

\bibitem{mockus2002two}
A.~Mockus, R.~T. Fielding, and J.~D. Herbsleb, ``Two case studies of open
  source software development: Apache and mozilla,'' \emph{TOSEM'02}, vol.~11,
  no.~3, pp. 309--346.

\bibitem{krishnamurthy2002cave}
S.~Krishnamurthy, ``Cave or community?: An empirical examination of 100 mature
  open source projects.''

\bibitem{robles2009evolution}
G.~Robles, J.~M. Gonzalez-Barahona, and I.~Herraiz, ``Evolution of the core
  team of developers in libre software projects,'' in \emph{MSR'09}, pp.
  167--170.

\bibitem{bird2009promises}
C.~Bird, P.~C. Rigby, E.~T. Barr, D.~J. Hamilton, D.~M. German, and P.~Devanbu,
  ``The promises and perils of mining git,'' in \emph{MSR'09}, pp. 1--10.

\bibitem{kalliamvakou2014promises}
E.~Kalliamvakou, G.~Gousios, K.~Blincoe, L.~Singer, D.~M. German, and
  D.~Damian, ``The promises and perils of mining github,'' in \emph{MSR'14},
  pp. 92--101.

\bibitem{joblin2015developer}
M.~Joblin, W.~Mauerer, S.~Apel, J.~Siegmund, and D.~Riehle, ``From developer
  networks to verified communities: a fine-grained approach,'' in
  \emph{ICSE'15}, pp. 563--573.

\bibitem{jermakovics2011mining}
A.~Jermakovics, A.~Sillitti, and G.~Succi, ``Mining and visualizing developer
  networks from version control systems,'' in \emph{CHASE'11}, pp. 24--31.

\bibitem{zhang2014developer}
W.~Zhang, L.~Nie, H.~Jiang, Z.~Chen, and J.~Liu, ``Developer social networks in
  software engineering: construction, analysis, and applications,''
  \emph{SCIS'14}, vol.~57, no.~12, pp. 1--23.

\bibitem{tymchuk2014collaboration}
Y.~Tymchuk, A.~Mocci, and M.~Lanza, ``Collaboration in open-source projects:
  Myth or reality?'' in \emph{MSR'14}, pp. 304--307.

\bibitem{joblin2017classifying}
M.~Joblin, S.~Apel, C.~Hunsen, and W.~Mauerer, ``Classifying developers into
  core and peripheral: An empirical study on count and network metrics,'' in
  \emph{ICSE'17}, pp. 164--174.

\bibitem{ccaglayan2016effect}
B.~{\c{C}}aglayan and A.~B. Bener, ``Effect of developer collaboration activity
  on software quality in two large scale projects,'' \emph{JSS'16}, vol. 118,
  pp. 288--296.

\bibitem{ren2018towards}
J.~Ren, H.~Yin, Q.~Hu, A.~Fox, and W.~Koszek, ``Towards quantifying the
  development value of code contributions,'' in \emph{FSE'18}, pp. 775--779.

\bibitem{palomba2018community}
F.~Palomba, D.~A. Tamburri, A.~Serebrenik, A.~Zaidman, F.~A. Fontana, and
  R.~Oliveto, ``How do community smells influence code smells?'' in
  \emph{ICSE-Companion'18}, pp. 240--241.

\bibitem{jermakovics2013exploring}
A.~Jermakovics, A.~Sillitti, and G.~Succi, ``Exploring collaboration networks
  in open-source projects,'' in \emph{IFIP-ICOS'13}, pp. 97--108.

\bibitem{joblin2017evolutionary}
M.~Joblin, S.~Apel, and W.~Mauerer, ``Evolutionary trends of developer
  coordination: A network approach,'' \emph{EMSE'17}, vol.~22, no.~4, pp.
  2050--2094.

\bibitem{pinzger2008can}
M.~Pinzger, N.~Nagappan, and B.~Murphy, ``Can developer-module networks predict
  failures?'' in \emph{FSE'08}, pp. 2--12.

\bibitem{thung2013network}
F.~Thung, T.~F. Bissyande, D.~Lo, and L.~Jiang, ``Network structure of social
  coding in github,'' in \emph{CSMR‘13}, pp. 323--326.

\bibitem{hong2011understanding}
Q.~Hong, S.~Kim, S.~Cheung, and C.~Bird, ``Understanding a developer social
  network and its evolution,'' in \emph{ICSM'11}, pp. 323--332.

\bibitem{bird2008latent}
C.~Bird, D.~Pattison, R.~D'Souza, V.~Filkov, and P.~Devanbu, ``Latent social
  structure in open source projects,'' in \emph{FSE'08}, pp. 24--35.

\bibitem{surian2010mining}
D.~Surian, D.~Lo, and E.-P. Lim, ``Mining collaboration patterns from a large
  developer network,'' in \emph{WCRE'10}, pp. 269--273.

\bibitem{wang2013devnet}
S.~Wang, W.~Zhang, Y.~Yang, and Q.~Wang, ``Devnet: exploring developer
  collaboration in heterogeneous networks of bug repositories,'' in
  \emph{ESEM'13}, pp. 193--202.

\bibitem{wolf2009mining}
T.~Wolf, A.~Schr{\"o}ter, D.~Damian, L.~D. Panjer, and T.~H. Nguyen, ``Mining
  task-based social networks to explore collaboration in software teams,''
  \emph{IEEE Software'09}, vol.~26, no.~1, pp. 58--66.

\bibitem{zhang2013heterogeneous}
W.~Zhang, S.~Wang, Y.~Yang, and Q.~Wang, ``Heterogeneous network analysis of
  developer contribution in bug repositories,'' in \emph{ICCSC'13}, pp.
  98--105.

\bibitem{surian2011recommending}
D.~Surian, N.~Liu, D.~Lo, H.~Tong, E.-P. Lim, and C.~Faloutsos, ``Recommending
  people in developers' collaboration network,'' in \emph{WCRE'11}, pp.
  379--388.

\bibitem{mcdonald2003recommending}
D.~W. McDonald, ``Recommending collaboration with social networks: a
  comparative evaluation,'' in \emph{CHI'03}, pp. 593--600.

\bibitem{jeong2009improving}
G.~Jeong, S.~Kim, and T.~Zimmermann, ``Improving bug triage with bug tossing
  graphs,'' in \emph{FSE'09}, pp. 111--120.

\bibitem{zanetti2013categorizing}
M.~S. Zanetti, I.~Scholtes, C.~J. Tessone, and F.~Schweitzer, ``Categorizing
  bugs with social networks: a case study on four open source software
  communities,'' in \emph{ICSE'13}, pp. 1032--1041.

\bibitem{bird2006mining}
C.~Bird, A.~Gourley, P.~Devanbu, M.~Gertz, and A.~Swaminathan, ``Mining email
  social networks,'' in \emph{MSR'06}, pp. 137--143.

\bibitem{zanetti2013rise}
M.~S. Zanetti, I.~Scholtes, C.~J. Tessone, and F.~Schweitzer, ``The rise and
  fall of a central contributor: dynamics of social organization and
  performance in the gentoo community,'' in \emph{CHASE'13}, pp. 49--56.

\bibitem{jiang2017mining}
H.~Jiang, J.~Zhang, H.~Ma, N.~Nazar, and Z.~Ren, ``Mining authorship
  characteristics in bug repositories,'' \emph{SCIS'17}, vol.~60, no.~1, p.
  012107.

\bibitem{zhou2015will}
M.~Zhou and A.~Mockus, ``Who will stay in the floss community? modeling
  participant’s initial behavior,'' \emph{TSE'15}, vol.~41, no.~1, pp.
  82--99.

\bibitem{zhou2012make}
------, ``What make long term contributors: Willingness and opportunity in oss
  community,'' in \emph{ICSE'12}, pp. 518--528.

\bibitem{gharehyazie2015developer}
M.~Gharehyazie, D.~Posnett, B.~Vasilescu, and V.~Filkov, ``Developer initiation
  and social interactions in oss: A case study of the apache software
  foundation,'' \emph{EMSE'15}, vol.~20, no.~5, pp. 1318--1353.

\bibitem{lopez2006applying}
L.~L{\'o}pez-Fern{\'a}ndez, G.~Robles, J.~M. Gonzalez-Barahona, and I.~Herraiz,
  ``Applying social network analysis techniques to community-driven libre
  software projects,'' \emph{IJITWE'06}, vol.~1, no.~3, pp. 27--48.

\bibitem{lopez2004applying}
L.~Lopez-Fernandez, G.~Robles, J.~M. Gonzalez-Barahona \emph{et~al.},
  ``Applying social network analysis to the information in cvs repositories,''
  in \emph{MSR'04}, p. 101–105.

\bibitem{toral2010analysis}
S.~L. Toral, M.~d.~R. Mart{\'\i}nez-Torres, and F.~Barrero, ``Analysis of
  virtual communities supporting oss projects using social network analysis,''
  \emph{IST'10}, vol.~52, no.~3, pp. 296--303.

\bibitem{canfora2011social}
G.~Canfora, L.~Cerulo, M.~Cimitile, and M.~Di~Penta, ``Social interactions
  around cross-system bug fixings: the case of freebsd and openbsd,'' in
  \emph{MSR'11}, pp. 143--152.

\bibitem{bird2011don}
C.~Bird, N.~Nagappan, B.~Murphy, H.~Gall, and P.~Devanbu, ``Don't touch my
  code!: examining the effects of ownership on software quality,'' in
  \emph{FSE'11}, pp. 4--14.

\bibitem{tsay2014influence}
J.~Tsay, L.~Dabbish, and J.~Herbsleb, ``Influence of social and technical
  factors for evaluating contribution in github,'' in \emph{ICSE'14}, pp.
  356--366.

\bibitem{rahman2011ownership}
F.~Rahman and P.~Devanbu, ``Ownership, experience and defects: a fine-grained
  study of authorship,'' in \emph{ICSE'11}, pp. 491--500.

\bibitem{zhou2015cross}
B.~Zhou, I.~Neamtiu, and R.~Gupta, ``A cross-platform analysis of bugs and
  bug-fixing in open source projects: Desktop vs. android vs. ios,'' in
  \emph{EASE'15}, p.~7.

\bibitem{izquierdo2011developers}
D.~Izquierdo-Cortazar, A.~Capiluppi, and J.~M. Gonzalez-Barahona, ``Are
  developers fixing their own bugs?: Tracing bug-fixing and bug-seeding
  committers,'' \emph{IJOSSP'11}, vol.~3, no.~2, pp. 23--42.

\bibitem{german2003gnome}
D.~M. German, ``The gnome project: a case study of open source, global software
  development,'' \emph{Software Process: Improvement and Practice}, vol.~8,
  no.~4, pp. 201--215, 2003.

\bibitem{xuan2012developer}
J.~Xuan, H.~Jiang, Z.~Ren, and W.~Zou, ``Developer prioritization in bug
  repositories,'' in \emph{ICSE'12}, pp. 25--35.

\bibitem{dinh2005freebsd}
T.~T. Dinh-Trong and J.~M. Bieman, ``The freebsd project: A replication case
  study of open source development,'' \emph{TSE'05}, vol.~31, no.~6, pp.
  481--494.

\bibitem{izurieta2006evolution}
C.~Izurieta and J.~Bieman, ``The evolution of freebsd and linux,'' in
  \emph{ISESE'06}, pp. 204--211.

\bibitem{tian2012identifying}
Y.~Tian, J.~Lawall, and D.~Lo, ``Identifying linux bug fixing patches,'' in
  \emph{ICSE'12}, pp. 386--396.

\bibitem{nvd}
U.~N.~I. of~Standards and Technology, ``National vulnerability database,''
  \url{https://nvd.nist.gov/home.cfm}.

\bibitem{cve}
M.~Corporation, ``Common vulnerabilities and exposures,''
  \url{https://cve.mitre.org/}.

\bibitem{kim2006automatic}
S.~Kim, T.~Zimmermann, K.~Pan, E.~James~Jr \emph{et~al.}, ``Automatic
  identification of bug-introducing changes,'' in \emph{ASE'06}, pp. 81--90.

\bibitem{wijayasekara2012mining}
D.~Wijayasekara, M.~Manic, J.~L. Wright, and M.~McQueen, ``Mining bug databases
  for unidentified software vulnerabilities,'' in \emph{ICHSI'12}, pp. 89--96.

\bibitem{ponta2019dataset}
S.~E. Ponta, H.~Plate, A.~Sabetta, M.~Bezzi, and C.~Dangremont, ``A
  manually-curated dataset of fixes to vulnerabilities of open-source
  software,'' in \emph{MSR'19}.

\bibitem{zhou2017automated}
Y.~Zhou and A.~Sharma, ``Automated identification of security issues from
  commit messages and bug reports,'' in \emph{FSE'17}, pp. 914--919.

\bibitem{SourceClear}
``Sourceclear,'' \url{https://www.sourceclear.com/}.

\bibitem{wang2019images}
J.~Wang, M.~Li, S.~Wang, T.~Menzies, and Q.~Wang, ``Images don’t lie:
  Duplicate crowdtesting reports detection with screenshot information,''
  \emph{IST'19}.

\bibitem{wang2016towards}
J.~Wang, Q.~Cui, Q.~Wang, and S.~Wang, ``Towards effectively test report
  classification to assist crowdsourced testing,'' in \emph{ESEM'16}, p.~6.

\bibitem{runeson2007detection}
P.~Runeson, M.~Alexandersson, and O.~Nyholm, ``Detection of duplicate defect
  reports using natural language processing,'' in \emph{ICSE'07}, pp. 499--510.

\bibitem{rocha2016empirical}
H.~Rocha, M.~T. Valente, H.~Marques-Neto, and G.~C. Murphy, ``An empirical
  study on recommendations of similar bugs,'' in \emph{SANER'16}, vol.~1, pp.
  46--56.

\bibitem{witten2016data}
I.~H. Witten, E.~Frank, M.~A. Hall, and C.~J. Pal, \emph{Data Mining: Practical
  machine learning tools and techniques}.\hskip 1em plus 0.5em minus
  0.4em\relax Morgan Kaufmann, 2016.

\bibitem{li2017large}
F.~Li and V.~Paxson, ``A large-scale empirical study of security patches,'' in
  \emph{CCS'17}, pp. 2201--2215.

\bibitem{perl2015vccfinder}
H.~Perl, S.~Dechand, M.~Smith, D.~Arp, F.~Yamaguchi, K.~Rieck, S.~Fahl, and
  Y.~Acar, ``Vccfinder: Finding potential vulnerabilities in open-source
  projects to assist code audits,'' in \emph{CCS'15}, pp. 426--437.

\bibitem{sliwerski2005changes}
J.~{\'S}liwerski, T.~Zimmermann, and A.~Zeller, ``When do changes induce
  fixes?'' in \emph{MSR'05}, vol.~30, no.~4, pp. 1--5.

\bibitem{da2017framework}
D.~A. da~Costa, S.~McIntosh, W.~Shang, U.~Kulesza, R.~Coelho, and A.~E. Hassan,
  ``A framework for evaluating the results of the szz approach for identifying
  bug-introducing changes,'' \emph{TSE'17}, vol.~43, no.~7, pp. 641--657.

\bibitem{jiang2013personalized}
T.~Jiang, L.~Tan, and S.~Kim, ``Personalized defect prediction,'' in
  \emph{ASE'13}, pp. 279--289.

\bibitem{gu2010has}
Z.~Gu, E.~T. Barr, D.~J. Hamilton, and Z.~Su, ``Has the bug really been
  fixed?'' in \emph{ICSE'10}, vol.~1, pp. 55--64.

\bibitem{robles2005developer}
G.~Robles and J.~M. Gonzalez-Barahona, ``Developer identification methods for
  integrated data from various sources,'' in \emph{MSR'05}, pp. 1--5.

\bibitem{ehrlich2012all}
K.~Ehrlich and M.~Cataldo, ``All-for-one and one-for-all?: a multi-level
  analysis of communication patterns and individual performance in
  geographically distributed software development,'' in \emph{CSCW'12}, pp.
  945--954.

\bibitem{giger2012can}
E.~Giger, M.~Pinzger, and H.~C. Gall, ``Can we predict types of code changes?
  an empirical analysis,'' in \emph{MSR'12}, pp. 217--226.

\bibitem{well2003research}
A.~D. Well and J.~L. Myers, \emph{Research design \& statistical
  analysis}.\hskip 1em plus 0.5em minus 0.4em\relax Psychology Press, 2003.

\bibitem{kim2011dealing}
S.~Kim, H.~Zhang, R.~Wu, and L.~Gong, ``Dealing with noise in defect
  prediction,'' in \emph{ICSE'11}, pp. 481--490.

\bibitem{astromskis2017patterns}
S.~Astromskis, G.~Bavota, A.~Janes, B.~Russo, and M.~Di~Penta, ``Patterns of
  developers behaviour: A 1000-hour industrial study,'' \emph{JSS'07}, vol.
  132, pp. 85--97.

\bibitem{bu2017program}
W.~Bu, M.~Xue, L.~Xu, Y.~Zhou, Z.~Tang, and T.~Xie, ``When program analysis
  meets mobile security: an industrial study of misusing android internet
  sockets,'' in \emph{FSE'17}, pp. 842--847.

\bibitem{munaiah2018assisted}
N.~Munaiah, ``Assisted discovery of software vulnerabilities,'' in
  \emph{ICSE'18}, pp. 464--467.

\bibitem{camilo2015bugs}
F.~Camilo, A.~Meneely, and M.~Nagappan, ``Do bugs foreshadow vulnerabilities?:
  a study of the chromium project,'' in \emph{MSR'15}, pp. 269--279.

\bibitem{decan2018impact}
A.~Decan, T.~Mens, and E.~Constantinou, ``On the impact of security
  vulnerabilities in the npm package dependency network,'' in \emph{MSR'18},
  pp. 181--191.

\bibitem{frei2006large}
S.~Frei, M.~May, U.~Fiedler, and B.~Plattner, ``Large-scale vulnerability
  analysis,'' in \emph{SIGCOMM'06}, pp. 131--138.

\bibitem{shahzad2012large}
M.~Shahzad, M.~Z. Shafiq, and A.~X. Liu, ``A large scale exploratory analysis
  of software vulnerability life cycles,'' in \emph{ICSE'12}, pp. 771--781.

\bibitem{zhong2015empirical}
H.~Zhong and Z.~Su, ``An empirical study on real bug fixes,'' in
  \emph{ICSE'15}, pp. 913--923.

\bibitem{mu2018understanding}
D.~Mu, A.~Cuevas, L.~Yang, H.~Hu, X.~Xing, B.~Mao, and G.~Wang, ``Understanding
  the reproducibility of crowd-reported security vulnerabilities,'' in
  \emph{USENIX Security'18)}, pp. 919--936.

\bibitem{ozment2006milk}
A.~Ozment and S.~E. Schechter, ``Milk or wine: does software security improve
  with age?'' in \emph{USENIX Security Symposium'06}.

\bibitem{xu2017spain}
Z.~Xu, B.~Chen, M.~Chandramohan, Y.~Liu, and F.~Song, ``Spain: security patch
  analysis for binaries towards understanding the pain and pills,'' in
  \emph{ICSE'17}, pp. 462--472.

\bibitem{walden2014predicting}
J.~Walden, J.~Stuckman, and R.~Scandariato, ``Predicting vulnerable components:
  Software metrics vs text mining,'' in \emph{ISSRE'14}, pp. 23--33.

\bibitem{medeiros2017software}
N.~Medeiros, N.~Ivaki, P.~Costa, and M.~Vieira, ``Software metrics as
  indicators of security vulnerabilities,'' in \emph{ISSRE'17}, pp. 216--227.

\bibitem{yang2016mhcp}
Y.~Yang, J.~Ai, X.~Li, and W.~E. Wong, ``Mhcp model for quality evaluation for
  software structure based on software complex network,'' in \emph{ISSRE'16},
  pp. 298--308.

\bibitem{park2012empirical}
J.~Park, M.~Kim, B.~Ray, and D.-H. Bae, ``An empirical study of supplementary
  bug fixes,'' in \emph{MSR'12}, pp. 40--49.

\end{thebibliography}
\end{document}